\documentclass[aps,showpacs,twocolumn,longbibliography]{revtex4-1}
\usepackage[utf8]{inputenc}
\usepackage{amsfonts}
\usepackage{amssymb}
\usepackage{amsmath}
\usepackage{graphicx}
\usepackage{epsfig}
\usepackage{subfigure}
\usepackage{appendix}
\usepackage{color}
\usepackage{hyperref}
\usepackage{clrscode3e}
\usepackage{fullpage}

\hypersetup{hypertex=true,
colorlinks=true,
linkcolor=blue,
anchorcolor=blue,
urlcolor=blue,
citecolor=blue}

\begin{document}

\title{Topological quantum slinky motion in resonant extended Bose-Hubbard
model}
\author{H. P. Zhang}
\author{Z. Song}
\email{songtc@nankai.edu.cn}

\begin{abstract}
We study the one-dimensional Bose-Hubbard model under the resonant
condition, where a series of quantum slinky oscillations occur in a two-site
system for boson numbers $n\in \lbrack 2,\infty )$. In the strong
interaction limit, it can be shown that the quantum slinky motions become
the dominant channels for boson propagation, which are described by a set of
effective non-interacting Hamiltonians. They are sets of generalized
Su-Schrieffer-Heeger chains with an $n$-site unit cell, referred to as
trimerization, tetramerization, and pentamerization, etc., possessing
non-trivial Zak phases. The corresponding edge states are demonstrated by
the $n$-boson bound states at the ends of the chains. We also investigate
the dynamic detection of edge boson clusters through an analysis of quench
dynamics. Numerical results indicate that stable edge oscillations clearly
manifest the interaction-induced topological features within the extended
Bose-Hubbard model.
\end{abstract}

\affiliation{School of Physics, Nankai University, Tianjin 300071, China}
\maketitle

\section{Introduction}

\label{Introduction} Topological theory has been well established in
condensed matter physics, since the discovery of an association between
integer quantum Hall conductance and topological Chern invariants \cite%
{Thouless1982}. The concepts of topology have been extensively studied in
both condensed matter physics and material sciences \cite%
{Kitaev2001,Ryu2002,Greiner2002,Murakami2004,Kane2005,Bernevig2006,Fu2007,Fu2007a,Schnyder2008,Ryu2010,Hasan2010,Qi2011,Xu2011,Burkov2011,Young2012,Wang2012,Wang2013,Bardyn2012,Tarruell2012,Lin2014,Weng2015,Lu2015,Leykam2016,Chiu2016,Kunst2018,Armitage2018}%
. One of the simplest models of topological insulators is the SSH model \cite%
{Su1979}, which describes particle hopping in a 1D lattice with staggered
hopping constants. This system supports localized zero-energy edge states
associated with non-trivial Zak phase \cite{Zak1989}.

So far, discussions of topological insulator models have primarily focused
on non-interacting systems. A natural question is whether the
particle-particle interaction can induce additional topological features?
Recently, it has been shown that an extended Bose-Hubbard model can support
topologically nontrivial edge and interface states of repulsively bound
pairs, even though it is topologically trivial in the single-particle regime
\cite{Stepanenko2020}. On the other hand, the dynamics of particle pairs in
lattice systems have garnered considerable interest, owing to the rapid
advancements in experimental techniques. Ultra-cold atoms have proven to be
an ideal testing ground for few-particle fundamental physics, as optical
lattices offer clean realizations of a variety of many-body Hamiltonians. It
stimulates many experimental \cite{Winkler2006,Foelling2007,Gustavsson2008}
and theoretical investigations \cite%
{Mahajan2006,Petrosyan2007,Creffield2007,Kuklov2007,Zoellner2008,Wang2008,Valiente2008,Jin2009,Valiente2009,Valiente2010,Javanainen2010,Wang2010,Rosch2008,Zhang2024}
in strongly correlated systems. The essential physics of the proposed bound
pair involves the periodic potential suppressing single-particle tunneling
across the barrier. This suppression prevents the decay of the pair, which
would otherwise occur. In contrast, there exists another type of bound pair
that permits correlated single-particle tunneling \cite{Jin2011,Lin2014}.
Such a bound pair acts as a quasi-particle, with an energy band width of the
same order as that of a single particle.

In this work, we extend the concept of such bound states to scenarios
involving a greater number of particles. We study the one-dimensional
Bose-Hubbard model under the resonant condition, where a series of quantum
slinky oscillations occur in a two-site system for boson numbers $n\in
\lbrack 2,\infty )$. A quantum slinky is analogues of a classical slinky,
which is a helical spring toy (see Fig. \ref{slinky_motion}). It offers a
method to achieve $n$-boson bound states within the Hubbard model.
Furthermore, it can be demonstrated that quantum slinky motions dominate the
channels for boson propagation, as described by a set of effective
non-interacting Hamiltonians in the strong interaction limit. They are sets
of generalized Su-Schrieffer-Heeger chains with an $n$-site unit cell,
referred to as trimerization, tetramerization, and pentamerization, etc.,
possessing non-trivial Zak phases. The corresponding edge states are
demonstrated by the $n$-boson bound states at the ends of the chains.
Numerical simulations are employed to demonstrate the dynamic detection of
edge boson clusters by analyzing quench dynamics. The numerical results
align with our predictions and clearly manifest the interaction-induced
topological features within the extended Bose-Hubbard model.

The remaining parts of this paper are organized as follows. Section \ref%
{Model and resonant dimer} describes the concept of quantum slinky\ in an
extended Bose-Hubbard model under resonant condition. Section \ref{Quantum
slinky mode}, gives the effective Hamiltonian and discusses its topological
properties under various boundary conditions. Section \ref{Edge slinky modes
and dynamic detections} is devoted to the static and dynamic detections of
edge slinky states. Finally, we give a summary and discussion in Section \ref%
{sec_Summary}.

\section{Model and resonant dimer}

\label{Model and resonant dimer} 
\begin{figure}[t]
\centering
\includegraphics[width=0.48\textwidth]{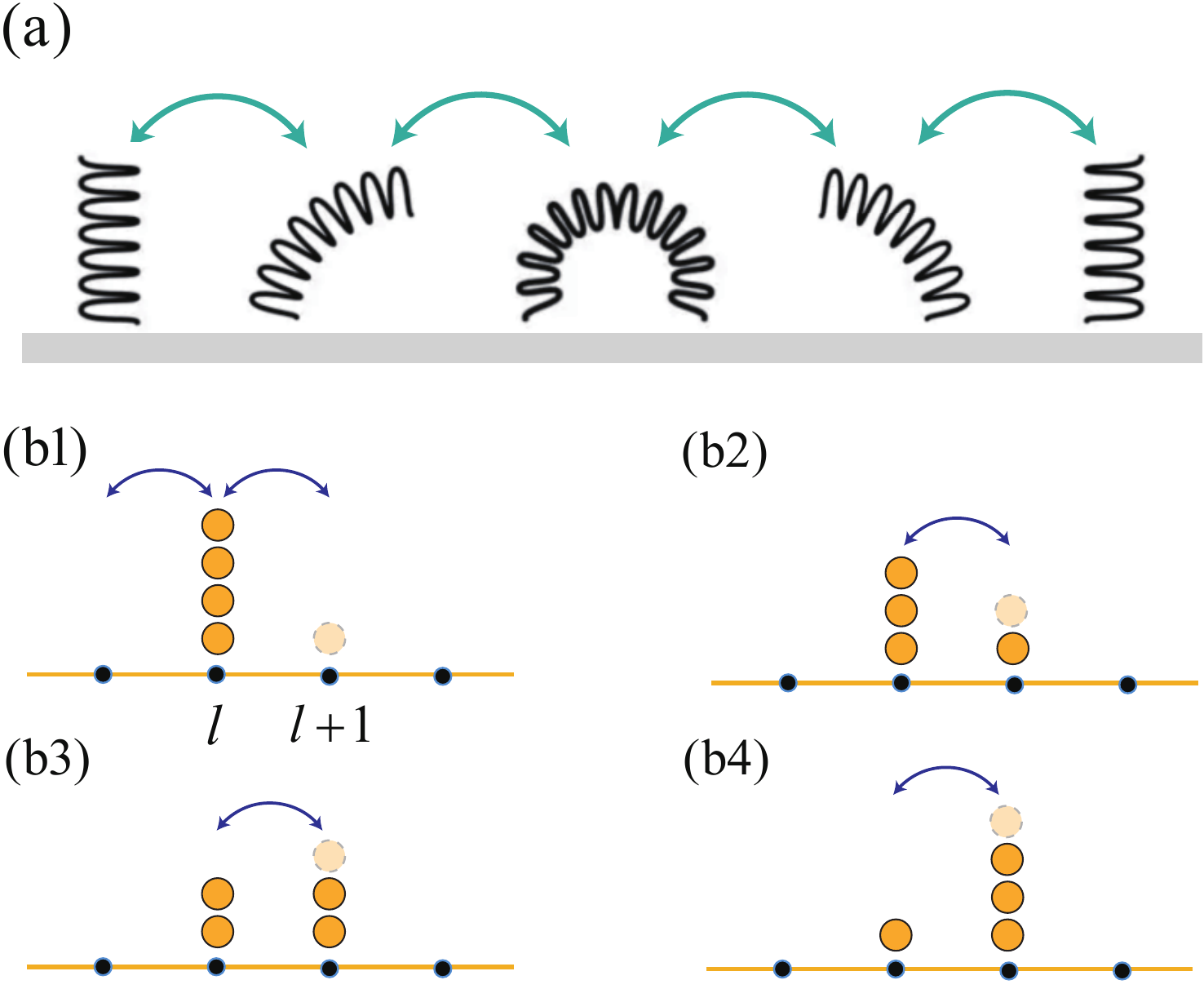}
\caption{Schematic illustrations of a classical and quantum slinky in the
extended Bose-Hubbard chain under resonant conditions. (a) The classical
slinky is a helical spring toy, which can move down a stair step by step.
Its dynamic feature is that it stretches and re-forms itself. (b1-b4) Four
configurations of the boson cluster with $n=4$. They are degenerate states
of the Hamiltonian $H_{\text{\textrm{V}}}$, which are analogues of a helical
spring in states of compression (b1) and tension (b2, b3, b4). The quantum
slinky can be driven by the quantum fluctuations of the term $H_{\text{ 
\textrm{T}}}$, while the classical slinky is driven by gravity and elastic
forces. The stable quantum slinky, which acts as a single particle within
the energy bands, can be realized in the strong interaction limit.
Furthermore, the edge-localized quantum slinky can be topologically
protected by the energy gap.}
\label{slinky_motion}
\end{figure}

We consider an extended Bose-Hubbard model describing interacting particles
in the lowest Bloch band of a one dimensional lattice, which can be employed
to describe ultracold atoms or molecules with magnetic or electric
dipole-dipole interactions in optical lattices. We focus on the dynamics of
the bosonic cluster, which is $n$ identical bosons in a bound state. For the
simplest case with $n=2$, it has been demonstrated that, as another type of
bound pair, it allows for correlated single-particle tunneling, as shown in
previous work \cite{Jin2011,Lin2014}. Such a bound pair can act as a
quasi-particle, with an energy band width of the same order as that of a
single particle. One of the objectives of this paper is to demonstrate that
this type of bound state can be generalized to an $n$-boson system.

We consider the Hamiltonian for one-dimensional extended Bose-Hubbard model
on a $N$-site lattice%
\begin{eqnarray}
H &=&H_{\mathrm{T}}+H_{\mathrm{V}},  \notag \\
H_{\mathrm{T}} &=&-\kappa \sum_{j=1}\left( a_{j}^{\dagger }a_{j+1}+\text{ 
\textrm{H.c}.}\right) ,  \notag \\
H_{\mathrm{V}} &=&\frac{U}{2}\sum_{j=1}n_{j}\left( n_{j}-1\right)
+V\sum_{j=1}n_{j}n_{j+1},
\end{eqnarray}%
where $a_{i}^{\dag }$ is the creation operator of the boson at the $i$th
site, the tunneling strength, on-site and NN interactions between bosons are
denoted by $\kappa $, $U$\ and $V$. There are many invariant subspaces for $%
H $ arising from the symmetries, for instance,%
\begin{equation}
\left[ \hat{n},H\right] =\left[ \hat{T}_{1},H\right] =0,
\end{equation}%
when the periodic boundary condition, $a_{j+N}=a_{j}$, is taken.\ Here two
operators are the total boson number operator%
\begin{equation}
\hat{n}=\sum_{j=1}^{N}a_{j}^{\dagger }a_{j},
\end{equation}%
and the translational operator defined as%
\begin{equation}
\hat{T}_{1}a_{j}\hat{T}_{1}^{-1}=a_{j+1}.
\end{equation}%
These simple features are useful for the following discussion.

In this work, we concentrate on the case with $U=V$, which is referred to as
the resonant condition. Let us begin by analyzing in detail the many-boson
problem within the Hamiltonian $H_{\mathrm{V}}$. Introducing boson-dimer
number operator

\begin{equation}
M_{j}=n_{j}+n_{j+1},
\end{equation}%
for two-site lattice $j$\ and $j+1$, the interaction terms can be written as%
\begin{eqnarray}
&&\frac{U}{2}[n_{j}\left( n_{j}-1\right) +n_{j+1}\left( n_{j+1}-1\right)
+2n_{j}n_{j+1}]  \notag \\
&=&\frac{U}{2}M_{j}(M_{j}-1).
\end{eqnarray}%
It is clear that both the inter- and intra-dimer interactions can be
considered as on-dimer interactions under the resonant condition.

In the invariant subspace with fixed boson number $n$, there is a set of
degenerate eigenstates $\left\{ \left\vert l\right\rangle ,l\in \lbrack
1,nN]\right\} $ of $H_{\mathrm{V}}$ with eigen energy $E_{\mathrm{V}}(n)$ $%
=Un(n-1)/2$, under the periodic boundary condition which are expressed as
the form%
\begin{equation}
\left\vert \left( j-1\right) n+\lambda \right\rangle =\frac{\left(
a_{j}^{\dagger }\right) ^{n+1-\lambda }\left( a_{j+1}^{\dagger }\right)
^{\lambda -1}}{\sqrt{\left( n+1-\lambda \right) !\left( \lambda -1\right) !}}
|\mathrm{vac}\rangle ,
\end{equation}%
with $j\in \lbrack 1,N]$\ and $\lambda \in \lbrack 1,n]$. Here $\left\vert 
\mathrm{vac}\right\rangle $\ is the vacuum state for the boson operator $%
a_{j}$.

We note that such a set of states has a special characteristic: all $n$
bosons occupy a single dimer, which acts as a bosonic cluster. In addition,
the nearest energy level next to $E_{\mathrm{V}}(n)$\ is $E_{\mathrm{V}%
}(n)-U $, resulting in an energy gap of $U$.\ These lead to an interesting
dynamic behavior when the hopping term $H_{\mathrm{T}}$\ is considered under
the strong correlation condition where $U\gg \kappa $. There is only one
channel for the transition from a dimer state to the next dimer state:%
\begin{equation}
H_{\mathrm{T}}\left\vert jn\right\rangle \rightarrow \left\vert
jn+1\right\rangle ,
\end{equation}%
or explicitly%
\begin{equation}
a_{j}^{\dagger }\left( a_{j+1}^{\dagger }\right) ^{n-1}|\mathrm{vac}\rangle
\rightarrow \left( a_{j+1}^{\dagger }\right) ^{n}|\mathrm{vac}\rangle .
\end{equation}%
where $\left\vert nN+1\right\rangle =\left\vert 1\right\rangle $. Then
quantum slinky motions become the dominant channels for boson propagation.
Such a constraint causes the dynamics of the bosonic cluster to move in a
manner akin to the slinky motion of a spring. In Fig. \ref{slinky_motion}, a
schematic illustration presents the analogies between the slinky motion of a
spring and the motion of a bosonic cluster with $n=4$.

\section{Topological edge modes}

\label{Quantum slinky mode}

\begin{figure*}[t]
\centering
\includegraphics[width=0.95\textwidth]{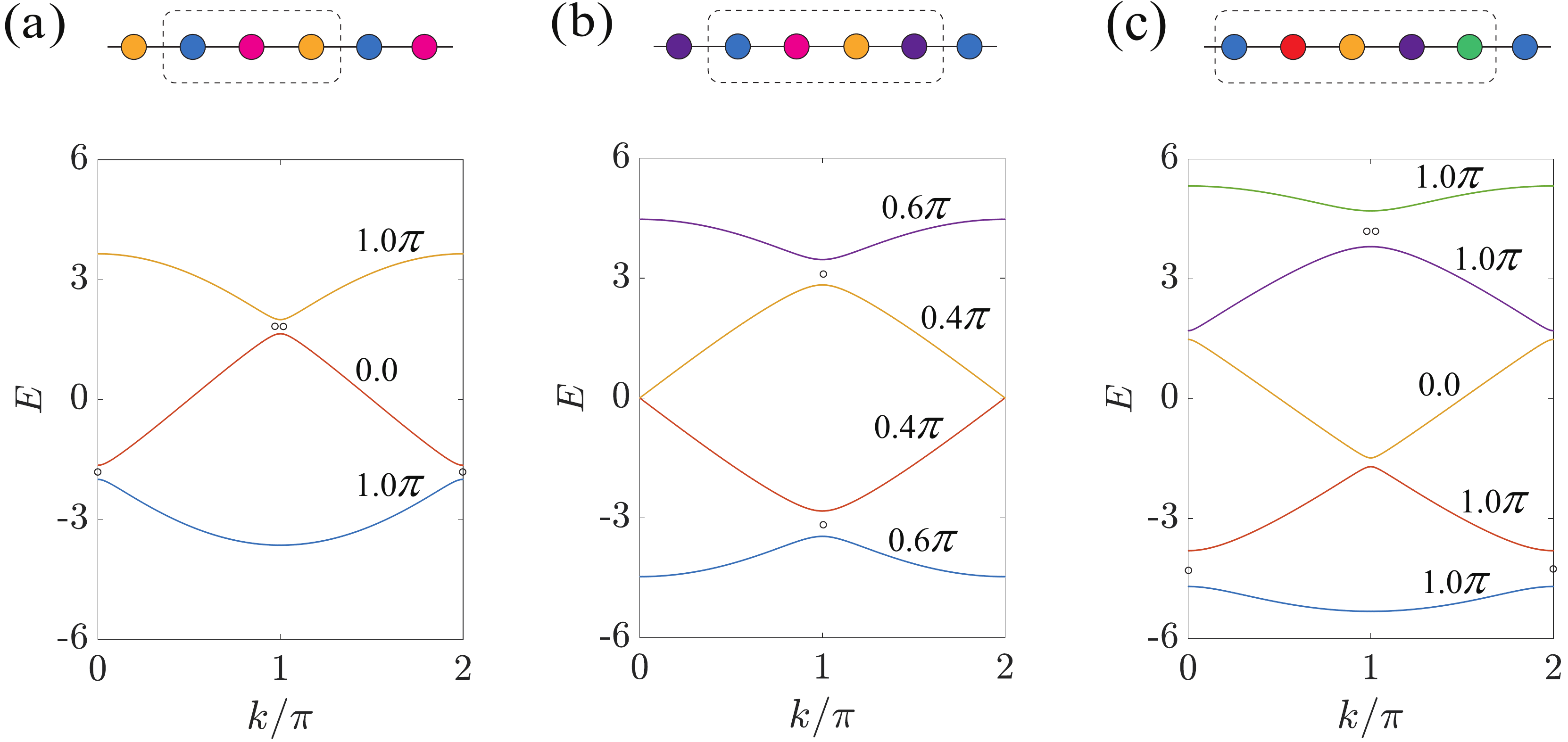}
\caption{Schematic illustrations of effective Hamiltonians in the strongly
correlation limit and the corresponding energy bands ($\protect\kappa =1$),
Zak phases, and edge states. (a, b, c) Schematics of the unit cells for the
effective Hamiltonians $H_{\mathrm{eff}}^{[3]}$, $H_{\mathrm{eff}}^{[4]}$,
and $H_{\mathrm{eff}}^{[5]}$, given in Eqs, (\protect\ref{Heff_3}), (\protect
\ref{Heff_4}), and (\protect\ref{Heff_5}),\ respectively. The energy bands
are obtained by exact diagonalization of the matrices in Eqs, (\protect\ref%
{non-trivial 3x3}), (\protect\ref{non-trivial 4x4}), and (\protect\ref%
{non-trivial 5x5}),\ respectively. The corresponding Zak phases of the bands
are indicated in the panels. The open boundary conditions are applied to the
Hamiltonians $H_{\mathrm{eff}}^{[3]}$, $H_{\mathrm{eff}}^{[4]}$, and $H_{%
\mathrm{eff}}^{[5]}$ by cutting off two neighboring unit cells. The nonzero
Zak phases are obtained from the matices in Eqs, (\protect\ref{non-trivial
3x3}), (\protect\ref{non-trivial 4x4}), and (\protect\ref{non-trivial 5x5}%
),\ respectively. When the open boundary conditions are matched to the three
matrices, there exist edge states, which are labeled by the black empty
circles within the gaps. It can be seen that in the cases of (a) and (c),
the Zak phases are nonzero and quantized, associated with pairs of edge
states, while in (b) only single edge states exist with non-quantized Zak
phases.}
\label{energy_band}
\end{figure*}
In the subspace spanned by the set of eigenstates $\left\{ \left\vert
l\right\rangle \right\} $, indicated by boson number $n$, the effective
Hamiltonian can be written as%
\begin{eqnarray}
H_{\mathrm{eff}}^{[n]} &=&-\kappa \sum_{j}h_{j}-\kappa \sum_{j}I_{j,j+1} 
\notag \\
&&+U\frac{n(n-1)}{2}\sum_{l=1}^{nN}\left\vert l\right\rangle \left\langle
l\right\vert ,
\end{eqnarray}%
where 
\begin{eqnarray}
h_{j}=&&\sum_{\lambda =1}^{n-1}\sqrt{(n-\lambda +1)\lambda }\left\vert
\left( j-1\right) n+\lambda \right\rangle  \notag \\
&&\times \left\langle \left( j-1\right) n+\lambda +1\right\vert +\text{ 
\textrm{H.c}.}, \\
I_{j,j+1}=&&\sqrt{n}\left\vert jn\right\rangle \left\langle jn+1\right\vert +%
\text{\textrm{H.c}.}.
\end{eqnarray}%
In its present form, $H_{\mathrm{eff}}^{[n]}$ are formally analogous to a
tight-binding model describing a single-particle dynamics in a ring with NN
hopping. It consists $N$ of unit cells denoted by the sub-Hamiltonian $%
\left\{ h_{j}\right\} $, on $n$-site lattice. Here $I_{j,j+1}$\ denotes the
hopping terms between two neighboring unit cells. In this paper, we are
interested in the dynamics of bosonic cluster, which corresponds to the
eigen problem of the effective Hamiltonian. {We would like to point out
that, the hopping strength is of the order of }$\kappa ${, not }$\kappa
^{2}/U${. This ensures that the phenomena arising from the effective
Hamiltonian can be observed in experiments, similar to those of a single
boson.}\ Based on the translational symmetry of the original system, the
effective Hamiltonian has the translational symmetry%
\begin{equation}
\hat{T}_{n}H_{\mathrm{eff}}^{[n]}\hat{T}_{n}^{-1}=H_{\mathrm{eff}}^{[n]},
\end{equation}%
where the translational operator is defined as%
\begin{equation}
\hat{T}_{n}\left\vert l\right\rangle =\left\vert l+n\right\rangle .
\end{equation}%
It indicates that $H_{\mathrm{eff}}^{[n]}$ can be solved by Fourier
transformation, and its spectrum consists of $n$ energy bands. The
corresponding complete set of eigenstates describe all the quantum slinky
modes. We would like to stress that for a given Hamiltonian $H_{\mathrm{eff}%
}^{[n]}$, the description of a unit cell is not unique due to the
translational symmetry. There are $n$ types of unit cell, each associated
with a different Fourier transformation. What is quite expected and
remarkable is that some energy bands may be topologically nontrivial,
meaning they possess a quantized Zak phase. Accordingly, a topologically
nontrivial Zak phase ensures the existence of topological edge slinky mode.
In the following, we consider several cases with small values of $n$. We
will present the explicit form of $H_{\mathrm{eff}}^{[n]}$, the Fourier
transformations for each type, the Zak phases of each energy band, and the
number of edge modes. The corresponding Hamiltonian $H_{\mathrm{eff}}^{[n]}$%
\ with open boundary conditions can be obtained by cutting off one of the
connections between two nearest neighboring unit cells. Each type of Fourier
transformation corresponds to a specific type of open chain. Consequently,
each set of Zak phases corresponds to a specific configuration of edge modes.

(i) In the case where $n=2$, the Hamiltonian is given by%
\begin{equation}
H_{\mathrm{eff}}^{[2]}=-\sqrt{2}\kappa \sum_{l=1}^{2N}(|l\rangle \langle
l+1|+\mathrm{H.c.}).
\end{equation}%
This represents a uniform chain that is topologically trivial. Here, we
neglect the uniform on-site potential.

(ii) In the case where $n=3$, the Hamiltonian is given by%
\begin{eqnarray}
H_{\mathrm{eff}}^{[3]} &=&-\kappa \sum_{j=1}^{N}(\sqrt{3}|3j-2\rangle
\langle 3j-1|+2|3j-1\rangle \langle 3j|  \notag \\
&&+\sqrt{3}|3j\rangle \langle 3j+1|+\mathrm{H.c.}).  \label{Heff_3}
\end{eqnarray}%
There are three types of Fourier transformations indexed by $\mu \in \lbrack
1,3]$, which can be expressed in the form%
\begin{equation}
|\mu ,k\rangle _{\nu }=\frac{1}{\sqrt{N}}\sum_{j=0}^{N-1}e^{-ikj}|3j+\mu
+\nu -1\rangle
\end{equation}%
with $\nu \in \lbrack 1,3]$ and $k=2\pi m/N$ ($m\in \lbrack 1,N]$). Then the
Hamiltonian $H_{\mathrm{eff}}^{[3]}$\ can be expressed as%
\begin{equation}
H_{\mathrm{eff}}^{[3]}=-\kappa \sum_{k}\sum_{\nu ,\nu ^{\prime }\in \lbrack
1,3]}|\mu ,k\rangle _{\nu }\left( h_{k}^{\mu }\right) _{\nu \nu ^{\prime
}}\langle \mu ,k|_{\nu ^{\prime }},
\end{equation}%
where three matrices $\left\{ h_{k}^{\mu }\right\} $\ are expressed
explicitly in the Appendix \ref{Appendix}. Here we only focus on one of them,%
\begin{equation}
h_{k}^{3}=\left( 
\begin{array}{ccc}
0 & \sqrt{3} & 2e^{-ik} \\ 
\sqrt{3} & 0 & \sqrt{3} \\ 
2e^{ik} & \sqrt{3} & 0%
\end{array}%
\right) ,  \label{non-trivial 3x3}
\end{equation}%
which is shown to be non-trivial from its eigenvectors under the periodic
and open boundary conditions. In Fig. \ref{energy_band}(a), the
corresponding energy band, Zak phases, and edge modes are presented. It
shows that the Zak phases are nonzero and quantized, associated with pairs
of mid-gap edge states.

(iii) In the case where $n=4$, the Hamiltonian is given by

\begin{eqnarray}
H_{\mathrm{eff}}^{[4]} &=&-\kappa \sum_{j=1}^{N}(2|4j-3\rangle \langle 4j-2|+%
\sqrt{6}|4j-2\rangle \langle 4j-1|\,  \notag \\
&&+\sqrt{6}|4j-1\rangle \langle 4j|\,+2|4j\rangle \langle 4j+1|+\mathrm{H.c.}%
).  \label{Heff_4}
\end{eqnarray}%
There are four types of Fourier transformations indexed by $\mu \in \lbrack
1,4]$, which can be expressed in the form%
\begin{equation}
|\mu ,k\rangle _{\nu }=\frac{1}{\sqrt{N}}\sum_{j=0}^{N-1}e^{-ikj}|4j+\mu
+\nu -1\rangle
\end{equation}%
with $\nu \in \lbrack 1,4]$ and $k=2\pi m/N$ ($m\in \lbrack 1,N]$). Then the
Hamiltonian $H_{\mathrm{eff}}^{[4]}$\ can be expressed as%
\begin{equation}
H_{\mathrm{eff}}^{[4]}=-\kappa \sum_{k}\sum_{\nu ,\nu ^{\prime }\in \lbrack
1,4]}|\mu ,k\rangle _{\nu }\left( h_{k}^{\mu }\right) _{\nu \nu ^{\prime
}}\langle \mu ,k|_{\nu ^{\prime }},
\end{equation}%
where four matrices $\left\{ h_{k}^{\mu }\right\} $\ are expressed
explicitly in the Appendix \ref{Appendix}. Here we only focus on one of them,%
\begin{equation}
h_{k}^{3}=\left( 
\begin{array}{cccc}
0 & \sqrt{6} & 0 & \sqrt{6}e^{-ik} \\ 
\sqrt{6} & 0 & 2 & 0 \\ 
0 & 2 & 0 & 2 \\ 
\sqrt{6}e^{ik} & 0 & 2 & 0%
\end{array}%
\right) ,  \label{non-trivial 4x4}
\end{equation}%
which is shown to be non-trivial from its eigenvectors under the periodic
and open boundary conditions. In Fig. \ref{energy_band}(b), the
corresponding energy band, Zak phases, and edge modes are presented. It
shows that single edge states exist, associated with non-quantized Zak
phases. These situations are similar to those in a Rice-Mele (RM) model \cite%
{Rice1982,Xiao2010,Wang2018,Wang2018a} with nonzero staggered on-site
potentials.

(iv) In the case where $n=5$, the Hamiltonian is given by%
\begin{eqnarray}
H_{\mathrm{eff}}^{[5]} &=&-\kappa \sum_{j=1}^{N}(\sqrt{5}|5j-4\rangle
\langle 5j-3|+2\sqrt{2}|5j-3\rangle \langle 5j-2|  \notag \\
&&+3|5j-2\rangle \langle 5j-1|+2\sqrt{2}|5j-1\rangle \langle 5j|  \notag \\
&&+\sqrt{5}|5j\rangle \langle 5j+1|+\mathrm{H.c.}).  \label{Heff_5}
\end{eqnarray}

There are five types of Fourier transformations indexed by $\mu \in \lbrack
1,5]$, which can be expressed in the form%
\begin{equation}
|\mu ,k\rangle _{\nu }=\frac{1}{\sqrt{N}}\sum_{j=0}^{N-1}e^{-ikj}|5j+\mu
+\nu -1\rangle
\end{equation}%
with $\nu \in \lbrack 1,5]$ and $k=2\pi m/N$ ($m\in \lbrack 1,N]$). Then the
Hamiltonian $H_{\mathrm{eff}}^{[5]}$\ can be expressed as 
\begin{equation}
H_{\mathrm{eff}}^{[5]}=-\kappa \sum_{k}\sum_{\nu ,\nu ^{\prime }\in \lbrack
1,5]}|\mu ,k\rangle _{\nu }\left( h_{k}^{\mu }\right) _{\nu \nu ^{\prime
}}\langle \mu ,k|_{\nu ^{\prime }},
\end{equation}%
where five matrices $\left\{ h_{k}^{\mu }\right\} $\ are expressed
explicitly in the Appendix \ref{Appendix}. Here we only focus on one of them,%
\begin{equation}
h_{k}^{4}=\left( 
\begin{array}{ccccc}
0 & 2\sqrt{2} & 0 & 0 & 3e^{-ik} \\ 
2\sqrt{2} & 0 & \sqrt{5} & 0 & 0 \\ 
0 & \sqrt{5} & 0 & \sqrt{5} & 0 \\ 
0 & 0 & \sqrt{5} & 0 & 2\sqrt{2} \\ 
3e^{ik} & 0 & 0 & 2\sqrt{2} & 0%
\end{array}%
\right) ,  \label{non-trivial 5x5}
\end{equation}%
which is shown to be non-trivial from its eigenvectors under the periodic
and open boundary conditions. In Fig. \ref{energy_band}(c), the
corresponding energy band, Zak phases, and edge modes are presented. It
shows that the Zak phases are nonzero and quantized, associated with pairs
of mid-gap edge states.

\section{Edge slinky modes and dynamic detections}

\label{Edge slinky modes and dynamic detections}

The analysis in the last section indicates that quantum slinky motions
become the dominant channels for boson propagation, which are described by a
set of effective non-interacting Hamiltonians in the strong interaction
limit. Some of these Hamiltonians are generalized Su-Schrieffer-Heeger (SSH)
chains with an $n$-site unit cell, referred to as trimerization,
tetramerization, and pentamerization, etc., possessing non-trivial Zak
phases and edge states.

Intuitively, such topological edge states can be observed in the original
boson systems. However, it should be noted that all types of $H_{\mathrm{eff}%
}^{[n]}$\ with open boundary conditions cannot be realized in an extended
Hubbard open chain, i.e., by cutting off the interactions between $1$st and $%
N$th sites. In order to realize edge slinky states in a resonant
Bose-Hubbard model, one can remove the basis $\left\vert l\right\rangle $,
which consists of the states $\left( a_{j-1}^{\dagger }\right) ^{n-k}\left(
a_{j}^{\dagger }\right) ^{k}|\mathrm{vac}\rangle $ and $\left(
a_{j}^{\dagger }\right) ^{k}\left( a_{j+1}^{\dagger }\right) ^{n-k}|\mathrm{%
vac}\rangle $ with $k>n_{0}$,\ by\ adding additional interactions of the form%
\begin{equation}
H_{\text{\textrm{imp}}}=W\prod_{\lambda \in \left[ 0,n_{0}\right]
}(a_{j}^{\dagger }a_{j}-\lambda ),
\end{equation}%
on the $1$st and $N$th sites with a very large $W$. Such impurities break
the translational symmetry,\ and can realize the target systems $H_{\mathrm{%
chain}}^{[n]}$ corresponding to the Hamiltonians $H_{\mathrm{eff}}^{[n]}$\
with open boundary conditions (the relationships between them are expressed
explicitly in the Appendix \ref{Appendix}). By this method, the topological
edge states can be demonstrated by the $n$-boson bound states at the ends of
the chains. To verify the above analysis, numerical simulations are
performed to investigate the static and dynamic detections of the edge boson
clusters.

Here we only focus on three typical cases, which correspond to three
matrices given in Eqs. (\ref{non-trivial 3x3}), (\ref{non-trivial 4x4}), and
(\ref{non-trivial 5x5}), respectively. Each of them corresponds to an
original Hubbard Hamiltonian $H_{\mathrm{chain}}^{[n]}(\mu )$ ($n=3,4,5;$ $%
\mu \in \lbrack 1,n]$), which is expected to exhibit edge states
corresponding to the nontrivial Zak phases of the energy bands of $H_{%
\mathrm{eff}}^{[n]}$. The corresponding Hamiltonians are given as following,

\begin{eqnarray}
H_{\mathrm{chain}}^{[3]}(3) &=&H+W\prod_{\lambda \in \left[ 0,1\right]
}(a_{1}^{\dagger }a_{1}-\lambda )  \notag \\
&&+W\prod_{\lambda ^{\prime }\in \left[ 0,1\right] }(a_{N}^{\dagger
}a_{N}-\lambda ^{\prime }),
\end{eqnarray}%
\begin{eqnarray}
H_{\mathrm{chain}}^{[4]}(3) &=&H+W\prod_{\lambda \in \left[ 0,2\right]
}(a_{1}^{\dagger }a_{1}-\lambda )  \notag \\
&&+W\prod_{\lambda ^{\prime }\in \left[ 0,1\right] }(a_{N}^{\dagger
}a_{N}-\lambda ^{\prime }),
\end{eqnarray}%
\begin{eqnarray}
H_{\mathrm{chain}}^{[5]}(4) &=&H+W\prod_{\lambda \in \left[ 0,2\right]
}(a_{1}^{\dagger }a_{1}-\lambda )  \notag \\
&&+W\prod_{\lambda ^{\prime }\in \left[ 0,2\right] }(a_{N}^{\dagger
}a_{N}-\lambda ^{\prime }),
\end{eqnarray}%
where $H$\ is the extended Hubbard open chain, i.e., by cutting off the
interactions between $1$st and $N$th sites.

\begin{figure}[t]
\centering
\includegraphics[width=0.49\textwidth]{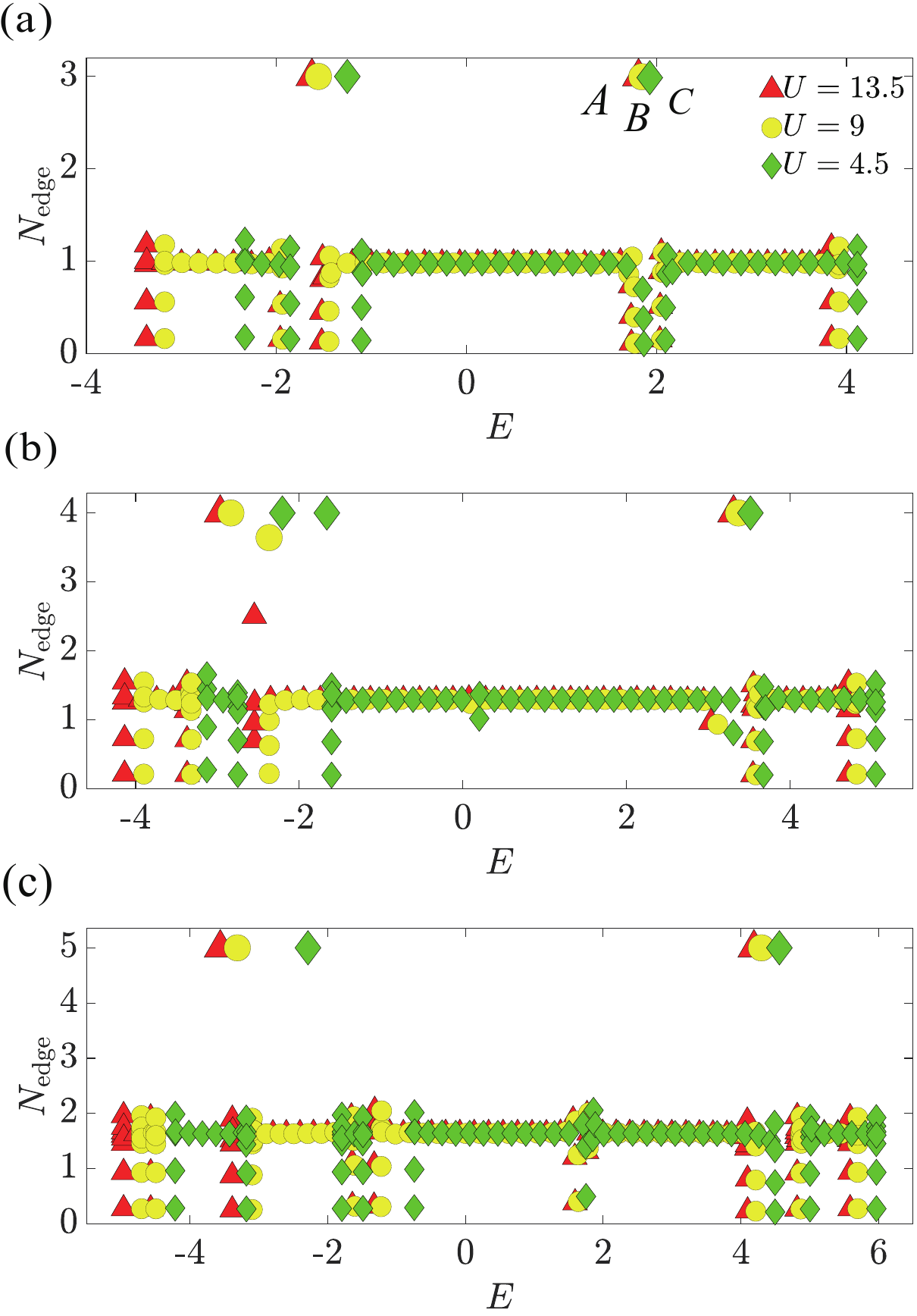}  
\caption{Plots of the edge boson number $N_{\mathrm{edge}}(E)$, as defined
by Eq.\ (\protect\ref{Nedge}), demonstrating the existence of the edge
slinky states. Panels (a), (b), and (c) correspond to the effective
Hamiltonians $H_{\mathrm{eff}}^{[3]}$, $H_{\mathrm{eff}}^{[4]}$, and $H_{ 
\mathrm{eff}}^{[5]}$, respectively, as shown in Eqs, (\protect\ref{Heff_3}),
(\protect\ref{Heff_4}), and (\protect\ref{Heff_5}). The numerical
simulations are performed under the truncation approximations. The plots of $%
N_{\mathrm{edge}}(E)$ do not include the datas for all the energy levels.
The system parameters for each panels are $N=200$, $W=30$, $\protect\kappa =1
$, $U=13.5$, $9.0$, and $4.5$, respectively. The results accord with our
predictions, even at moderate values of $U$.\ Three edge states, labeled A,
B, and C, in panel (a), are selected as the initial states for the time
evolutions shown in Fig. \protect\ref{oscillation}(a), (b), and (c),
respectively.}
\label{edge_partical}
\end{figure}

To measure the edge boson clusters, we introduce a quantity, the edge boson
number, denoted by $N_{\mathrm{edge}}(E)$, which is given by:%
\begin{equation}
N_{\mathrm{edge}}(E)=\sum_{j\in \text{\textrm{edge region}}}\left\langle
\psi \right\vert a_{j}^{\dag }a_{j}\left\vert \psi \right\rangle ,
\label{Nedge}
\end{equation}%
for the eigenstate $\left\vert \psi \right\rangle $ with eigenenergy $E$. We
compute $\left\vert \psi \right\rangle $\ and $E$\ for finite-size lattice
with fixed $n=3$, $4$, and $5$ by numerically diagonalizing the Hamiltonians 
$H_{\mathrm{chain}}^{[3]}(3)$, $H_{\mathrm{chain}}^{[4]}(3)$, and $H_{%
\mathrm{chain}}^{[5]}(4)$.\ In Fig. \ref{edge_partical}, quantities $N_{%
\mathrm{edge}}(E)$ are plotted for several cases with typical values of $U$.
Here, we employ a truncation approximation by selecting a set of basis
around the slinky states. Consequently, and the involved eigenstates $%
\left\vert \psi \right\rangle $\ do not constitute a complete set.\ As
expected, the peaks of $N_{\mathrm{edge}}(E)$ are evident near the position
of the energy gaps, particularly in the cases with large $U$. These peaks
take values near $3.0$, $4.0$, and $5.0$, respectively. Notably, this holds
true even at moderate values of $U$. Based on these observations, one can
investigate the dynamic signature of the edge slinky states.

\begin{figure*}[t]
\centering
\includegraphics[width=1\textwidth]{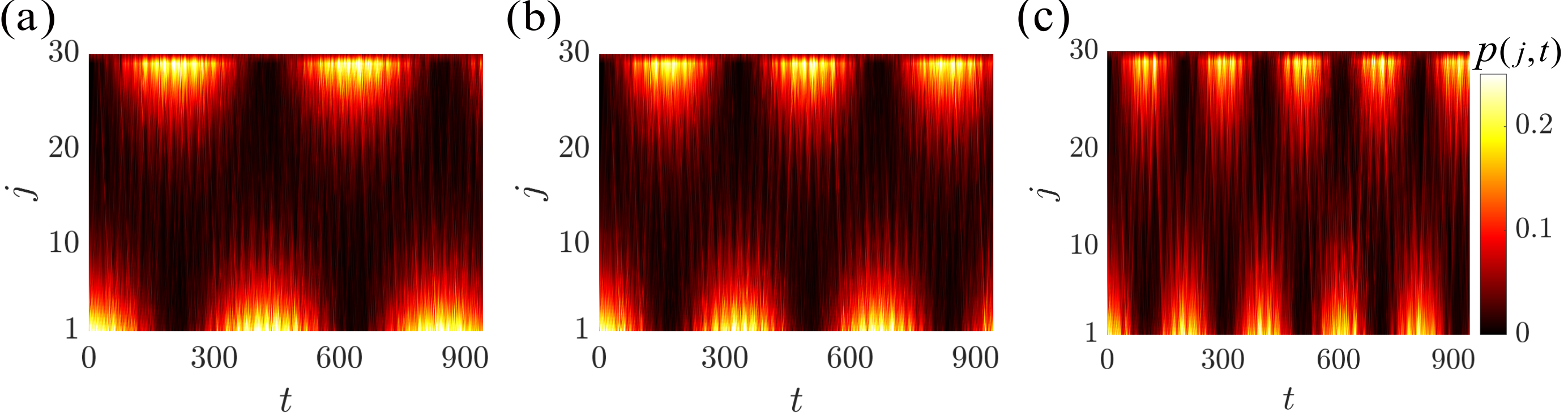}
\caption{Plots of the particle number distribution $p(j,t)$, as defined in
Eq.\ (\protect\ref{p(jt)}), demonstrating the quench dynamics of the edge
slinky states. Panels (a), (b), and (c) correspond to the cases where $W=30$%
, $\protect\kappa =1$, $U=13.5$, $9.0$, and $4.5$, respectively. The size of
the quench Hamiltonian $H_{\text{quen}}$ is $30$. The initial states for the
time evolutions shown in panels (a), (b), and (c) correspond to the edge
states labeled A, B, and C, respectively, as indicated in Fig.\ \protect\ref%
{edge_partical}(a). In each case, the system exhibits evident two-level
oscillations. As $U$ decreases, the frequency of oscillation increases,
which suggests a direct relationship between the energy gap and the value of 
$U$. This provides a dynamic signature indicative of the existence of edge
slinky states, which are induced by resonant Hubbard interactions.}
\label{oscillation}
\end{figure*}

In fact, we can detect the existence of such edge boson clusters by
analyzing quench dynamics. We simulate numerically the dynamic detection of
edge boson clusters by computing a quench process: The initial state $%
\left\vert \Psi \left( 0\right) \right\rangle $ is an edge state at one end
of the Hamiltonian $H_{\mathrm{chain}}^{[3]}(3)$\ on a larger lattice, while
the quench Hamiltonian $H_{\text{quen}}$ is also $H_{\mathrm{chain}%
}^{[3]}(3) $\ but on a small-size system. For a small-sized system, two
degenerate edge states can become two hybridized states separated by a small
energy gap. We plot the particle number distribution profile of $\left\vert
\Psi \left( t\right) \right\rangle $\ 
\begin{equation}
p(j,t)=\left\langle \Psi \left( t\right) \right\vert a_{j}^{\dag
}a_{j}\left\vert \Psi \left( t\right) \right\rangle =\left\vert a_{j}e^{-iH_{%
\text{quen}}t}\left\vert \Psi \left( 0\right) \right\rangle \right\vert ^{2}
\label{p(jt)}
\end{equation}%
for several cases with typical values of $U$ in Fig. \ref{oscillation}. As
predicted, we observe stable oscillations with a single frequency in each
case. This indicates that the oscillation frequency increases as the value
of $U$ decreases. Such behavior provides a dynamic signature indicative of
the existence of edge slinky states, which are induced by resonant Hubbard
interactions.

\section{Summary}

\label{sec_Summary}

In summary, the coherent dynamics of $n$ correlated bosons in a
one-dimensional extended Hubbard model with identical on-site $U$ and
nearest-neighbor site $V$ interactions, have been theoretically
investigated. The analysis reveals that in the resonant case, there always
exists an $n$-boson slinky state, which has a comparable bandwidth to that
of a single boson. In the strong interaction limit, it has been shown that
quantum slinky motions can be described by a set of generalized
Su-Schrieffer-Heeger chains with an $n$-site unit cell. Accordingly, when
the energy band possess non-trivial Zak phases, the corresponding edge
states appear as $n$-boson bound states at the ends of the chains. This
ensures that the topological $n$-boson bound states can be observed in
experiments, similar to those of a single boson. To this end, a dynamic
detection of edge boson clusters through an analysis of quench dynamics is
proposed. The stable edge oscillations predicted in this paper are an
exclusive signature of multi-boson clusters, as they are indicative of the
interaction-induced topological features within the extended Bose-Hubbard
model.

\acknowledgments This work was supported by National Natural Science
Foundation of China (under Grant No. 12374461).

\section*{Appendix}

\label{Appendix}

In this appendix, we present the explicit forms of the set of matices $%
\left\{ h_{k}^{\mu }\right\} $\ ($\mu \in \lbrack 1,n]$) obtained by the
effective Hamiltonian $H_{\mathrm{eff}}^{[n]}$, where $n=3$, $4$, and $5$.
Each $h_{k}^{\mu }$\ corresponds to an original Hubbard Hamiltonian $H_{%
\mathrm{chain}}^{[n]}(\mu )$, which is expected to exhibit edge states
corresponding to the nontrivial Zak phases of the energy bands of $H_{%
\mathrm{eff}}^{[n]}$.

(i) For the case with $n=3$, the set of matrices $h_{k}^{\mu }$ are
expressed explicitly as

\begin{eqnarray}
&&h_{k}^{1,2,3}=\left( 
\begin{array}{ccc}
0 & \sqrt{3} & \sqrt{3}e^{-ik} \\ 
\sqrt{3} & 0 & 2 \\ 
\sqrt{3}e^{ik} & 2 & 0%
\end{array}%
\right) , \\
&&\left( 
\begin{array}{ccc}
0 & 2 & \sqrt{3}e^{-ik} \\ 
2 & 0 & \sqrt{3} \\ 
\sqrt{3}e^{ik} & \sqrt{3} & 0%
\end{array}%
\right) ,\left( 
\begin{array}{ccc}
0 & \sqrt{3} & 2e^{-ik} \\ 
\sqrt{3} & 0 & \sqrt{3} \\ 
2e^{ik} & \sqrt{3} & 0%
\end{array}%
\right) .  \notag
\end{eqnarray}%
On the other hand, by adding the impurities, we obtain the explicit forms of$%
\ H_{\mathrm{chain}}^{[n]}(\mu )$, which are%
\begin{equation}
H_{\mathrm{chain}}^{[3]}(1)=H+W\prod_{\lambda ^{\prime }\in \left[ 0,2\right]
}(a_{N}^{\dagger }a_{N}-\lambda ^{\prime }),
\end{equation}%
\begin{equation}
H_{\mathrm{chain}}^{[3]}(2)=H+W\prod_{\lambda \in \left[ 0,2\right]
}(a_{1}^{\dagger }a_{1}-\lambda ),
\end{equation}%
\begin{eqnarray}
H_{\mathrm{chain}}^{[3]}(3) &=&H+W\prod_{\lambda \in \left[ 0,1\right]
}(a_{1}^{\dagger }a_{1}-\lambda )  \notag \\
&&+W\prod_{\lambda ^{\prime }\in \left[ 0,1\right] }(a_{N}^{\dagger
}a_{N}-\lambda ^{\prime }),  \label{Hchain3}
\end{eqnarray}%
under the condition $W\gg U$.

(ii) For the case with $n=4$, the set of matrices $h_{k}^{\mu }$ are
expressed explicitly as

\begin{equation}
h_{k}^{1}=\left( 
\begin{array}{cccc}
0 & 2 & 0 & 2e^{-ik} \\ 
2 & 0 & \sqrt{6} & 0 \\ 
0 & \sqrt{6} & 0 & \sqrt{6} \\ 
2e^{ik} & 0 & \sqrt{6} & 0%
\end{array}%
\right) ,
\end{equation}%
\begin{equation}
h_{k}^{2}=\left( 
\begin{array}{cccc}
0 & \sqrt{6} & 0 & 2e^{-ik} \\ 
\sqrt{6} & 0 & \sqrt{6} & 0 \\ 
0 & \sqrt{6} & 0 & 2 \\ 
2e^{ik} & 0 & 2 & 0%
\end{array}%
\right) ,
\end{equation}%
\begin{equation}
h_{k}^{3}=\left( 
\begin{array}{cccc}
0 & \sqrt{6} & 0 & \sqrt{6}e^{-ik} \\ 
\sqrt{6} & 0 & 2 & 0 \\ 
0 & 2 & 0 & 2 \\ 
\sqrt{6}e^{ik} & 0 & 2 & 0%
\end{array}%
\right) ,
\end{equation}%
\begin{equation}
h_{k}^{4}=\left( 
\begin{array}{cccc}
0 & 2 & 0 & \sqrt{6}e^{-ik} \\ 
2 & 0 & 2 & 0 \\ 
0 & 2 & 0 & \sqrt{6} \\ 
\sqrt{6}e^{ik} & 0 & \sqrt{6} & 0%
\end{array}%
\right) .
\end{equation}%
On the other hand, by adding the impurities, we obtain the explicit forms of$%
\ H_{\mathrm{chain}}^{[n]}(\mu )$, which are%
\begin{equation}
H_{\mathrm{chain}}^{[4]}(1)=H+W\prod_{\lambda ^{\prime }\in \left[ 0,3\right]
}(a_{N}^{\dagger }a_{N}-\lambda ^{\prime }),
\end{equation}%
\begin{equation}
H_{\mathrm{chain}}^{[4]}(2)=H+W\prod_{\lambda \in \left[ 0,3\right]
}(a_{1}^{\dagger }a_{1}-\lambda ),
\end{equation}%
\begin{eqnarray}
H_{\mathrm{chain}}^{[4]}(3) &=&H+W\prod_{\lambda \in \left[ 0,2\right]
}(a_{1}^{\dagger }a_{1}-\lambda )  \notag \\
&&+W\prod_{\lambda ^{\prime }\in \left[ 0,1\right] }(a_{N}^{\dagger
}a_{N}-\lambda ^{\prime }),  \label{Hchain4}
\end{eqnarray}%
\begin{eqnarray}
H_{\mathrm{chain}}^{[4]}(4) &=&H+W\prod_{\lambda \in \left[ 0,1\right]
}(a_{1}^{\dagger }a_{1}-\lambda ) \\
&&+W\prod_{\lambda ^{\prime }\in \left[ 0,2\right] }(a_{N}^{\dagger
}a_{N}-\lambda ^{\prime }),
\end{eqnarray}%
under the condition $W\gg U$.

(iii) For the case with $n=5$, the set of matrices $h_{k}^{\mu }$ are
expressed explicitly as%
\begin{equation}
h_{k}^{1}=\left( 
\begin{array}{ccccc}
0 & \sqrt{5} & 0 & 0 & \sqrt{5}e^{-ik} \\ 
\sqrt{5} & 0 & 2\sqrt{2} & 0 & 0 \\ 
0 & 2\sqrt{2} & 0 & 3 & 0 \\ 
0 & 0 & 3 & 0 & 2\sqrt{2} \\ 
\sqrt{5}e^{ik} & 0 & 0 & 2\sqrt{2} & 0%
\end{array}%
\right) ,
\end{equation}%
\begin{equation}
h_{k}^{2}=\left( 
\begin{array}{ccccc}
0 & 2\sqrt{2} & 0 & 0 & \sqrt{5}e^{-ik} \\ 
2\sqrt{2} & 0 & 3 & 0 & 0 \\ 
0 & 3 & 0 & 2\sqrt{2} & 0 \\ 
0 & 0 & 2\sqrt{2} & 0 & \sqrt{5} \\ 
\sqrt{5}e^{ik} & 0 & 0 & \sqrt{5} & 0%
\end{array}%
\right) ,
\end{equation}%
\begin{equation}
h_{k}^{3}=\left( 
\begin{array}{ccccc}
0 & 3 & 0 & 0 & 2\sqrt{2}e^{-ik} \\ 
3 & 0 & 2\sqrt{2} & 0 & 0 \\ 
0 & 2\sqrt{2} & 0 & \sqrt{5} & 0 \\ 
0 & 0 & \sqrt{5} & 0 & \sqrt{5} \\ 
2\sqrt{2}e^{ik} & 0 & 0 & \sqrt{5} & 0%
\end{array}%
\right) ,
\end{equation}%
\begin{equation}
h_{k}^{4}=\left( 
\begin{array}{ccccc}
0 & 2\sqrt{2} & 0 & 0 & 3e^{-ik} \\ 
2\sqrt{2} & 0 & \sqrt{5} & 0 & 0 \\ 
0 & \sqrt{5} & 0 & \sqrt{5} & 0 \\ 
0 & 0 & \sqrt{5} & 0 & 2\sqrt{2} \\ 
3e^{ik} & 0 & 0 & 2\sqrt{2} & 0%
\end{array}%
\right) ,
\end{equation}%
\begin{equation}
h_{k}^{5}=\left( 
\begin{array}{ccccc}
0 & \sqrt{5} & 0 & 0 & 2\sqrt{2}e^{-ik} \\ 
\sqrt{5} & 0 & \sqrt{5} & 0 & 0 \\ 
0 & \sqrt{5} & 0 & 2\sqrt{2} & 0 \\ 
0 & 0 & 2\sqrt{2} & 0 & 3 \\ 
2\sqrt{2}e^{ik} & 0 & 0 & 3 & 0%
\end{array}%
\right) .
\end{equation}%
On the other hand, by adding the impurities, we obtain the explicit forms of$%
\ H_{\mathrm{chain}}^{[5]}(\mu )$, which are%
\begin{equation}
H_{\mathrm{chain}}^{[5]}(1)=H+W\prod_{\lambda ^{\prime }\in \left[ 0,4\right]
}(a_{N}^{\dagger }a_{N}-\lambda ^{\prime }),
\end{equation}%
\begin{equation}
H_{\mathrm{chain}}^{[5]}(2)=H+W\prod_{\lambda \in \left[ 0,4\right]
}(a_{1}^{\dagger }a_{1}-\lambda ),
\end{equation}%
\begin{eqnarray}
H_{\mathrm{chain}}^{[5]}(3) &=&H+W\prod_{\lambda \in \left[ 0,3\right]
}(a_{1}^{\dagger }a_{1}-\lambda )  \notag \\
&&+W\prod_{\lambda ^{\prime }\in \left[ 0,1\right] }(a_{N}^{\dagger
}a_{N}-\lambda ^{\prime }),
\end{eqnarray}%
\begin{eqnarray}
H_{\mathrm{chain}}^{[5]}(4) &=&H+W\prod_{\lambda \in \left[ 0,2\right]
}(a_{1}^{\dagger }a_{1}-\lambda )  \notag \\
&&+W\prod_{\lambda ^{\prime }\in \left[ 0,2\right] }(a_{N}^{\dagger
}a_{N}-\lambda ^{\prime }),  \label{Hchain5}
\end{eqnarray}%
\begin{eqnarray}
H_{\mathrm{chain}}^{[5]}(5) &=&H+W\prod_{\lambda \in \left[ 0,1\right]
}(a_{1}^{\dagger }a_{1}-\lambda )  \notag \\
&&+W\prod_{\lambda ^{\prime }\in \left[ 0,3\right] }(a_{N}^{\dagger
}a_{N}-\lambda ^{\prime }),
\end{eqnarray}%
under the condition $W\gg U$.


\begin{thebibliography}{53}%
	\makeatletter
	\providecommand \@ifxundefined [1]{%
		\@ifx{#1\undefined}
	}%
	\providecommand \@ifnum [1]{%
		\ifnum #1\expandafter \@firstoftwo
		\else \expandafter \@secondoftwo
		\fi
	}%
	\providecommand \@ifx [1]{%
		\ifx #1\expandafter \@firstoftwo
		\else \expandafter \@secondoftwo
		\fi
	}%
	\providecommand \natexlab [1]{#1}%
	\providecommand \enquote  [1]{``#1''}%
	\providecommand \bibnamefont  [1]{#1}%
	\providecommand \bibfnamefont [1]{#1}%
	\providecommand \citenamefont [1]{#1}%
	\providecommand \href@noop [0]{\@secondoftwo}%
	\providecommand \href [0]{\begingroup \@sanitize@url \@href}%
	\providecommand \@href[1]{\@@startlink{#1}\@@href}%
	\providecommand \@@href[1]{\endgroup#1\@@endlink}%
	\providecommand \@sanitize@url [0]{\catcode `\\12\catcode `\$12\catcode
		`\&12\catcode `\#12\catcode `\^12\catcode `\_12\catcode `\%12\relax}%
	\providecommand \@@startlink[1]{}%
	\providecommand \@@endlink[0]{}%
	\providecommand \url  [0]{\begingroup\@sanitize@url \@url }%
	\providecommand \@url [1]{\endgroup\@href {#1}{\urlprefix }}%
	\providecommand \urlprefix  [0]{URL }%
	\providecommand \Eprint [0]{\href }%
	\providecommand \doibase [0]{http://dx.doi.org/}%
	\providecommand \selectlanguage [0]{\@gobble}%
	\providecommand \bibinfo  [0]{\@secondoftwo}%
	\providecommand \bibfield  [0]{\@secondoftwo}%
	\providecommand \translation [1]{[#1]}%
	\providecommand \BibitemOpen [0]{}%
	\providecommand \bibitemStop [0]{}%
	\providecommand \bibitemNoStop [0]{.\EOS\space}%
	\providecommand \EOS [0]{\spacefactor3000\relax}%
	\providecommand \BibitemShut  [1]{\csname bibitem#1\endcsname}%
	\let\auto@bib@innerbib\@empty
	\bibitem [{\citenamefont {Thouless}\ \emph {et~al.}(1982)\citenamefont
		{Thouless}, \citenamefont {Kohmoto}, \citenamefont {Nightingale},\ and\
		\citenamefont {den Nijs}}]{Thouless1982}%
	\BibitemOpen
	\bibfield  {author} {\bibinfo {author} {\bibfnamefont {D.~J.}\ \bibnamefont
			{Thouless}}, \bibinfo {author} {\bibfnamefont {M.}~\bibnamefont {Kohmoto}},
		\bibinfo {author} {\bibfnamefont {M.~P.}\ \bibnamefont {Nightingale}}, \ and\
		\bibinfo {author} {\bibfnamefont {M.}~\bibnamefont {den Nijs}},\ }\bibfield
	{title} {\enquote {\bibinfo {title} {Quantized hall conductance in a
				two-dimensional periodic potential},}\ }\href {\doibase
		10.1103/physrevlett.49.405} {\bibfield  {journal} {\bibinfo  {journal}
			{Physical Review Letters}\ }\textbf {\bibinfo {volume} {49}},\ \bibinfo
		{pages} {405--408} (\bibinfo {year} {1982})}\BibitemShut {NoStop}%
	\bibitem [{\citenamefont {Kitaev}(2001)}]{Kitaev2001}%
	\BibitemOpen
	\bibfield  {author} {\bibinfo {author} {\bibfnamefont {A~Yu}\ \bibnamefont
			{Kitaev}},\ }\bibfield  {title} {\enquote {\bibinfo {title} {Unpaired
				majorana fermions in quantum wires},}\ }\href {\doibase
		10.1070/1063-7869/44/10s/s29} {\bibfield  {journal} {\bibinfo  {journal}
			{Physics-Uspekhi}\ }\textbf {\bibinfo {volume} {44}},\ \bibinfo {pages}
		{131--136} (\bibinfo {year} {2001})}\BibitemShut {NoStop}%
	\bibitem [{\citenamefont {Ryu}\ and\ \citenamefont {Hatsugai}(2002)}]{Ryu2002}%
	\BibitemOpen
	\bibfield  {author} {\bibinfo {author} {\bibfnamefont {Shinsei}\ \bibnamefont
			{Ryu}}\ and\ \bibinfo {author} {\bibfnamefont {Yasuhiro}\ \bibnamefont
			{Hatsugai}},\ }\bibfield  {title} {\enquote {\bibinfo {title} {Topological
				origin of zero-energy edge states in particle-hole symmetric systems},}\
	}\href {\doibase 10.1103/physrevlett.89.077002} {\bibfield  {journal}
		{\bibinfo  {journal} {Physical Review Letters}\ }\textbf {\bibinfo {volume}
			{89}},\ \bibinfo {pages} {077002} (\bibinfo {year} {2002})}\BibitemShut
	{NoStop}%
	\bibitem [{\citenamefont {Greiner}\ \emph {et~al.}(2002)\citenamefont
		{Greiner}, \citenamefont {Mandel}, \citenamefont {Esslinger}, \citenamefont
		{Hänsch},\ and\ \citenamefont {Bloch}}]{Greiner2002}%
	\BibitemOpen
	\bibfield  {author} {\bibinfo {author} {\bibfnamefont {Markus}\ \bibnamefont
			{Greiner}}, \bibinfo {author} {\bibfnamefont {Olaf}\ \bibnamefont {Mandel}},
		\bibinfo {author} {\bibfnamefont {Tilman}\ \bibnamefont {Esslinger}},
		\bibinfo {author} {\bibfnamefont {Theodor~W.}\ \bibnamefont {Hänsch}}, \
		and\ \bibinfo {author} {\bibfnamefont {Immanuel}\ \bibnamefont {Bloch}},\
	}\bibfield  {title} {\enquote {\bibinfo {title} {Quantum phase transition
				from a superfluid to a mott insulator in a gas of ultracold atoms},}\ }\href
	{\doibase 10.1038/415039a} {\bibfield  {journal} {\bibinfo  {journal}
			{Nature}\ }\textbf {\bibinfo {volume} {415}},\ \bibinfo {pages} {39--44}
		(\bibinfo {year} {2002})}\BibitemShut {NoStop}%
	\bibitem [{\citenamefont {Murakami}\ \emph {et~al.}(2004)\citenamefont
		{Murakami}, \citenamefont {Nagaosa},\ and\ \citenamefont
		{Zhang}}]{Murakami2004}%
	\BibitemOpen
	\bibfield  {author} {\bibinfo {author} {\bibfnamefont {Shuichi}\ \bibnamefont
			{Murakami}}, \bibinfo {author} {\bibfnamefont {Naoto}\ \bibnamefont
			{Nagaosa}}, \ and\ \bibinfo {author} {\bibfnamefont {Shou-Cheng}\
			\bibnamefont {Zhang}},\ }\bibfield  {title} {\enquote {\bibinfo {title}
			{Spin-hall insulator},}\ }\href {\doibase 10.1103/physrevlett.93.156804}
	{\bibfield  {journal} {\bibinfo  {journal} {Physical Review Letters}\
		}\textbf {\bibinfo {volume} {93}},\ \bibinfo {pages} {156804} (\bibinfo
		{year} {2004})}\BibitemShut {NoStop}%
	\bibitem [{\citenamefont {Kane}\ and\ \citenamefont {Mele}(2005)}]{Kane2005}%
	\BibitemOpen
	\bibfield  {author} {\bibinfo {author} {\bibfnamefont {C.~L.}\ \bibnamefont
			{Kane}}\ and\ \bibinfo {author} {\bibfnamefont {E.~J.}\ \bibnamefont
			{Mele}},\ }\bibfield  {title} {\enquote {\bibinfo {title} {Quantum spin hall
				effect in graphene},}\ }\href {\doibase 10.1103/physrevlett.95.226801}
	{\bibfield  {journal} {\bibinfo  {journal} {Physical Review Letters}\
		}\textbf {\bibinfo {volume} {95}},\ \bibinfo {pages} {226801} (\bibinfo
		{year} {2005})}\BibitemShut {NoStop}%
	\bibitem [{\citenamefont {Bernevig}\ \emph {et~al.}(2006)\citenamefont
		{Bernevig}, \citenamefont {Hughes},\ and\ \citenamefont
		{Zhang}}]{Bernevig2006}%
	\BibitemOpen
	\bibfield  {author} {\bibinfo {author} {\bibfnamefont {B.~Andrei}\
			\bibnamefont {Bernevig}}, \bibinfo {author} {\bibfnamefont {Taylor~L.}\
			\bibnamefont {Hughes}}, \ and\ \bibinfo {author} {\bibfnamefont {Shou-Cheng}\
			\bibnamefont {Zhang}},\ }\bibfield  {title} {\enquote {\bibinfo {title}
			{Quantum spin hall effect and topological phase transition in hgte quantum
				wells},}\ }\href {\doibase 10.1126/science.1133734} {\bibfield  {journal}
		{\bibinfo  {journal} {Science}\ }\textbf {\bibinfo {volume} {314}},\ \bibinfo
		{pages} {1757--1761} (\bibinfo {year} {2006})}\BibitemShut {NoStop}%
	\bibitem [{\citenamefont {Fu}\ and\ \citenamefont {Kane}(2007)}]{Fu2007}%
	\BibitemOpen
	\bibfield  {author} {\bibinfo {author} {\bibfnamefont {Liang}\ \bibnamefont
			{Fu}}\ and\ \bibinfo {author} {\bibfnamefont {C.~L.}\ \bibnamefont {Kane}},\
	}\bibfield  {title} {\enquote {\bibinfo {title} {Topological insulators with
				inversion symmetry},}\ }\href {\doibase 10.1103/physrevb.76.045302}
	{\bibfield  {journal} {\bibinfo  {journal} {Physical Review B}\ }\textbf
		{\bibinfo {volume} {76}},\ \bibinfo {pages} {045302} (\bibinfo {year}
		{2007})}\BibitemShut {NoStop}%
	\bibitem [{\citenamefont {Fu}\ \emph {et~al.}(2007)\citenamefont {Fu},
		\citenamefont {Kane},\ and\ \citenamefont {Mele}}]{Fu2007a}%
	\BibitemOpen
	\bibfield  {author} {\bibinfo {author} {\bibfnamefont {Liang}\ \bibnamefont
			{Fu}}, \bibinfo {author} {\bibfnamefont {C.~L.}\ \bibnamefont {Kane}}, \ and\
		\bibinfo {author} {\bibfnamefont {E.~J.}\ \bibnamefont {Mele}},\ }\bibfield
	{title} {\enquote {\bibinfo {title} {Topological insulators in three
				dimensions},}\ }\href {\doibase 10.1103/physrevlett.98.106803} {\bibfield
		{journal} {\bibinfo  {journal} {Physical Review Letters}\ }\textbf {\bibinfo
			{volume} {98}},\ \bibinfo {pages} {106803} (\bibinfo {year}
		{2007})}\BibitemShut {NoStop}%
	\bibitem [{\citenamefont {Schnyder}\ \emph {et~al.}(2008)\citenamefont
		{Schnyder}, \citenamefont {Ryu}, \citenamefont {Furusaki},\ and\
		\citenamefont {Ludwig}}]{Schnyder2008}%
	\BibitemOpen
	\bibfield  {author} {\bibinfo {author} {\bibfnamefont {Andreas~P.}\
			\bibnamefont {Schnyder}}, \bibinfo {author} {\bibfnamefont {Shinsei}\
			\bibnamefont {Ryu}}, \bibinfo {author} {\bibfnamefont {Akira}\ \bibnamefont
			{Furusaki}}, \ and\ \bibinfo {author} {\bibfnamefont {Andreas W.~W.}\
			\bibnamefont {Ludwig}},\ }\bibfield  {title} {\enquote {\bibinfo {title}
			{Classification of topological insulators and superconductors in three
				spatial dimensions},}\ }\href {\doibase 10.1103/physrevb.78.195125}
	{\bibfield  {journal} {\bibinfo  {journal} {Physical Review B}\ }\textbf
		{\bibinfo {volume} {78}},\ \bibinfo {pages} {195125} (\bibinfo {year}
		{2008})}\BibitemShut {NoStop}%
	\bibitem [{\citenamefont {Ryu}\ \emph {et~al.}(2010)\citenamefont {Ryu},
		\citenamefont {Schnyder}, \citenamefont {Furusaki},\ and\ \citenamefont
		{Ludwig}}]{Ryu2010}%
	\BibitemOpen
	\bibfield  {author} {\bibinfo {author} {\bibfnamefont {Shinsei}\ \bibnamefont
			{Ryu}}, \bibinfo {author} {\bibfnamefont {Andreas~P}\ \bibnamefont
			{Schnyder}}, \bibinfo {author} {\bibfnamefont {Akira}\ \bibnamefont
			{Furusaki}}, \ and\ \bibinfo {author} {\bibfnamefont {Andreas W~W}\
			\bibnamefont {Ludwig}},\ }\bibfield  {title} {\enquote {\bibinfo {title}
			{Topological insulators and superconductors: tenfold way and dimensional
				hierarchy},}\ }\href {\doibase 10.1088/1367-2630/12/6/065010} {\bibfield
		{journal} {\bibinfo  {journal} {New Journal of Physics}\ }\textbf {\bibinfo
			{volume} {12}},\ \bibinfo {pages} {065010} (\bibinfo {year}
		{2010})}\BibitemShut {NoStop}%
	\bibitem [{\citenamefont {Hasan}\ and\ \citenamefont {Kane}(2010)}]{Hasan2010}%
	\BibitemOpen
	\bibfield  {author} {\bibinfo {author} {\bibfnamefont {M.~Z.}\ \bibnamefont
			{Hasan}}\ and\ \bibinfo {author} {\bibfnamefont {C.~L.}\ \bibnamefont
			{Kane}},\ }\bibfield  {title} {\enquote {\bibinfo {title} {Colloquium:
				Topological insulators},}\ }\href {\doibase 10.1103/revmodphys.82.3045}
	{\bibfield  {journal} {\bibinfo  {journal} {Reviews of Modern Physics}\
		}\textbf {\bibinfo {volume} {82}},\ \bibinfo {pages} {3045--3067} (\bibinfo
		{year} {2010})}\BibitemShut {NoStop}%
	\bibitem [{\citenamefont {Qi}\ and\ \citenamefont {Zhang}(2011)}]{Qi2011}%
	\BibitemOpen
	\bibfield  {author} {\bibinfo {author} {\bibfnamefont {Xiao-Liang}\
			\bibnamefont {Qi}}\ and\ \bibinfo {author} {\bibfnamefont {Shou-Cheng}\
			\bibnamefont {Zhang}},\ }\bibfield  {title} {\enquote {\bibinfo {title}
			{Topological insulators and superconductors},}\ }\href {\doibase
		10.1103/revmodphys.83.1057} {\bibfield  {journal} {\bibinfo  {journal}
			{Reviews of Modern Physics}\ }\textbf {\bibinfo {volume} {83}},\ \bibinfo
		{pages} {1057--1110} (\bibinfo {year} {2011})}\BibitemShut {NoStop}%
	\bibitem [{\citenamefont {Xu}\ \emph {et~al.}(2011)\citenamefont {Xu},
		\citenamefont {Weng}, \citenamefont {Wang}, \citenamefont {Dai},\ and\
		\citenamefont {Fang}}]{Xu2011}%
	\BibitemOpen
	\bibfield  {author} {\bibinfo {author} {\bibfnamefont {Gang}\ \bibnamefont
			{Xu}}, \bibinfo {author} {\bibfnamefont {Hongming}\ \bibnamefont {Weng}},
		\bibinfo {author} {\bibfnamefont {Zhijun}\ \bibnamefont {Wang}}, \bibinfo
		{author} {\bibfnamefont {Xi}~\bibnamefont {Dai}}, \ and\ \bibinfo {author}
		{\bibfnamefont {Zhong}\ \bibnamefont {Fang}},\ }\bibfield  {title} {\enquote
		{\bibinfo {title} {Chern semimetal and the quantized anomalous hall effect
				inhgcr2se4},}\ }\href {\doibase 10.1103/physrevlett.107.186806} {\bibfield
		{journal} {\bibinfo  {journal} {Physical Review Letters}\ }\textbf {\bibinfo
			{volume} {107}},\ \bibinfo {pages} {186806} (\bibinfo {year}
		{2011})}\BibitemShut {NoStop}%
	\bibitem [{\citenamefont {Burkov}\ and\ \citenamefont
		{Balents}(2011)}]{Burkov2011}%
	\BibitemOpen
	\bibfield  {author} {\bibinfo {author} {\bibfnamefont {A.~A.}\ \bibnamefont
			{Burkov}}\ and\ \bibinfo {author} {\bibfnamefont {Leon}\ \bibnamefont
			{Balents}},\ }\bibfield  {title} {\enquote {\bibinfo {title} {Weyl semimetal
				in a topological insulator multilayer},}\ }\href {\doibase
		10.1103/physrevlett.107.127205} {\bibfield  {journal} {\bibinfo  {journal}
			{Physical Review Letters}\ }\textbf {\bibinfo {volume} {107}},\ \bibinfo
		{pages} {127205} (\bibinfo {year} {2011})}\BibitemShut {NoStop}%
	\bibitem [{\citenamefont {Young}\ \emph {et~al.}(2012)\citenamefont {Young},
		\citenamefont {Zaheer}, \citenamefont {Teo}, \citenamefont {Kane},
		\citenamefont {Mele},\ and\ \citenamefont {Rappe}}]{Young2012}%
	\BibitemOpen
	\bibfield  {author} {\bibinfo {author} {\bibfnamefont {S.~M.}\ \bibnamefont
			{Young}}, \bibinfo {author} {\bibfnamefont {S.}~\bibnamefont {Zaheer}},
		\bibinfo {author} {\bibfnamefont {J.~C.~Y.}\ \bibnamefont {Teo}}, \bibinfo
		{author} {\bibfnamefont {C.~L.}\ \bibnamefont {Kane}}, \bibinfo {author}
		{\bibfnamefont {E.~J.}\ \bibnamefont {Mele}}, \ and\ \bibinfo {author}
		{\bibfnamefont {A.~M.}\ \bibnamefont {Rappe}},\ }\bibfield  {title} {\enquote
		{\bibinfo {title} {Dirac semimetal in three dimensions},}\ }\href {\doibase
		10.1103/physrevlett.108.140405} {\bibfield  {journal} {\bibinfo  {journal}
			{Physical Review Letters}\ }\textbf {\bibinfo {volume} {108}},\ \bibinfo
		{pages} {140405} (\bibinfo {year} {2012})}\BibitemShut {NoStop}%
	\bibitem [{\citenamefont {Wang}\ \emph {et~al.}(2012)\citenamefont {Wang},
		\citenamefont {Sun}, \citenamefont {Chen}, \citenamefont {Franchini},
		\citenamefont {Xu}, \citenamefont {Weng}, \citenamefont {Dai},\ and\
		\citenamefont {Fang}}]{Wang2012}%
	\BibitemOpen
	\bibfield  {author} {\bibinfo {author} {\bibfnamefont {Zhijun}\ \bibnamefont
			{Wang}}, \bibinfo {author} {\bibfnamefont {Yan}\ \bibnamefont {Sun}},
		\bibinfo {author} {\bibfnamefont {Xing-Qiu}\ \bibnamefont {Chen}}, \bibinfo
		{author} {\bibfnamefont {Cesare}\ \bibnamefont {Franchini}}, \bibinfo
		{author} {\bibfnamefont {Gang}\ \bibnamefont {Xu}}, \bibinfo {author}
		{\bibfnamefont {Hongming}\ \bibnamefont {Weng}}, \bibinfo {author}
		{\bibfnamefont {Xi}~\bibnamefont {Dai}}, \ and\ \bibinfo {author}
		{\bibfnamefont {Zhong}\ \bibnamefont {Fang}},\ }\bibfield  {title} {\enquote
		{\bibinfo {title} {Dirac semimetal and topological phase transitions ina3bi
				(a=na, k, rb)},}\ }\href {\doibase 10.1103/physrevb.85.195320} {\bibfield
		{journal} {\bibinfo  {journal} {Physical Review B}\ }\textbf {\bibinfo
			{volume} {85}},\ \bibinfo {pages} {195320} (\bibinfo {year}
		{2012})}\BibitemShut {NoStop}%
	\bibitem [{\citenamefont {Wang}\ \emph {et~al.}(2013)\citenamefont {Wang},
		\citenamefont {Weng}, \citenamefont {Wu}, \citenamefont {Dai},\ and\
		\citenamefont {Fang}}]{Wang2013}%
	\BibitemOpen
	\bibfield  {author} {\bibinfo {author} {\bibfnamefont {Zhijun}\ \bibnamefont
			{Wang}}, \bibinfo {author} {\bibfnamefont {Hongming}\ \bibnamefont {Weng}},
		\bibinfo {author} {\bibfnamefont {Quansheng}\ \bibnamefont {Wu}}, \bibinfo
		{author} {\bibfnamefont {Xi}~\bibnamefont {Dai}}, \ and\ \bibinfo {author}
		{\bibfnamefont {Zhong}\ \bibnamefont {Fang}},\ }\bibfield  {title} {\enquote
		{\bibinfo {title} {Three-dimensional dirac semimetal and quantum transport in
				cd3as2},}\ }\href {\doibase 10.1103/physrevb.88.125427} {\bibfield  {journal}
		{\bibinfo  {journal} {Physical Review B}\ }\textbf {\bibinfo {volume} {88}},\
		\bibinfo {pages} {125427} (\bibinfo {year} {2013})}\BibitemShut {NoStop}%
	\bibitem [{\citenamefont {Bardyn}\ \emph {et~al.}(2012)\citenamefont {Bardyn},
		\citenamefont {Baranov}, \citenamefont {Rico}, \citenamefont {İmamoğlu},
		\citenamefont {Zoller},\ and\ \citenamefont {Diehl}}]{Bardyn2012}%
	\BibitemOpen
	\bibfield  {author} {\bibinfo {author} {\bibfnamefont {C.-E.}\ \bibnamefont
			{Bardyn}}, \bibinfo {author} {\bibfnamefont {M.~A.}\ \bibnamefont {Baranov}},
		\bibinfo {author} {\bibfnamefont {E.}~\bibnamefont {Rico}}, \bibinfo {author}
		{\bibfnamefont {A.}~\bibnamefont {İmamoğlu}}, \bibinfo {author}
		{\bibfnamefont {P.}~\bibnamefont {Zoller}}, \ and\ \bibinfo {author}
		{\bibfnamefont {S.}~\bibnamefont {Diehl}},\ }\bibfield  {title} {\enquote
		{\bibinfo {title} {Majorana modes in driven-dissipative atomic superfluids
				with a zero chern number},}\ }\href {\doibase 10.1103/physrevlett.109.130402}
	{\bibfield  {journal} {\bibinfo  {journal} {Physical Review Letters}\
		}\textbf {\bibinfo {volume} {109}},\ \bibinfo {pages} {130402} (\bibinfo
		{year} {2012})}\BibitemShut {NoStop}%
	\bibitem [{\citenamefont {Tarruell}\ \emph {et~al.}(2012)\citenamefont
		{Tarruell}, \citenamefont {Greif}, \citenamefont {Uehlinger}, \citenamefont
		{Jotzu},\ and\ \citenamefont {Esslinger}}]{Tarruell2012}%
	\BibitemOpen
	\bibfield  {author} {\bibinfo {author} {\bibfnamefont {Leticia}\ \bibnamefont
			{Tarruell}}, \bibinfo {author} {\bibfnamefont {Daniel}\ \bibnamefont
			{Greif}}, \bibinfo {author} {\bibfnamefont {Thomas}\ \bibnamefont
			{Uehlinger}}, \bibinfo {author} {\bibfnamefont {Gregor}\ \bibnamefont
			{Jotzu}}, \ and\ \bibinfo {author} {\bibfnamefont {Tilman}\ \bibnamefont
			{Esslinger}},\ }\bibfield  {title} {\enquote {\bibinfo {title} {Creating,
				moving and merging dirac points with a fermi gas in a tunable honeycomb
				lattice},}\ }\href {\doibase 10.1038/nature10871} {\bibfield  {journal}
		{\bibinfo  {journal} {Nature}\ }\textbf {\bibinfo {volume} {483}},\ \bibinfo
		{pages} {302--305} (\bibinfo {year} {2012})}\BibitemShut {NoStop}%
	\bibitem [{\citenamefont {Lin}\ \emph {et~al.}(2014)\citenamefont {Lin},
		\citenamefont {Zhang},\ and\ \citenamefont {Song}}]{Lin2014}%
	\BibitemOpen
	\bibfield  {author} {\bibinfo {author} {\bibfnamefont {S.}~\bibnamefont
			{Lin}}, \bibinfo {author} {\bibfnamefont {X.~Z.}\ \bibnamefont {Zhang}}, \
		and\ \bibinfo {author} {\bibfnamefont {Z.}~\bibnamefont {Song}},\ }\bibfield
	{title} {\enquote {\bibinfo {title} {Sudden death of particle-pair bloch
				oscillation and unidirectional propagation in a one-dimensional driven
				optical lattice},}\ }\href {\doibase 10.1103/physreva.90.063411} {\bibfield
		{journal} {\bibinfo  {journal} {Physical Review A}\ }\textbf {\bibinfo
			{volume} {90}},\ \bibinfo {pages} {063411} (\bibinfo {year}
		{2014})}\BibitemShut {NoStop}%
	\bibitem [{\citenamefont {Weng}\ \emph {et~al.}(2015)\citenamefont {Weng},
		\citenamefont {Fang}, \citenamefont {Fang}, \citenamefont {Bernevig},\ and\
		\citenamefont {Dai}}]{Weng2015}%
	\BibitemOpen
	\bibfield  {author} {\bibinfo {author} {\bibfnamefont {Hongming}\
			\bibnamefont {Weng}}, \bibinfo {author} {\bibfnamefont {Chen}\ \bibnamefont
			{Fang}}, \bibinfo {author} {\bibfnamefont {Zhong}\ \bibnamefont {Fang}},
		\bibinfo {author} {\bibfnamefont {B.~Andrei}\ \bibnamefont {Bernevig}}, \
		and\ \bibinfo {author} {\bibfnamefont {Xi}~\bibnamefont {Dai}},\ }\bibfield
	{title} {\enquote {\bibinfo {title} {Weyl semimetal phase in
				noncentrosymmetric transition-metal monophosphides},}\ }\href {\doibase
		10.1103/physrevx.5.011029} {\bibfield  {journal} {\bibinfo  {journal}
			{Physical Review X}\ }\textbf {\bibinfo {volume} {5}},\ \bibinfo {pages}
		{011029} (\bibinfo {year} {2015})}\BibitemShut {NoStop}%
	\bibitem [{\citenamefont {Lu}\ \emph {et~al.}(2015)\citenamefont {Lu},
		\citenamefont {Wang}, \citenamefont {Ye}, \citenamefont {Ran}, \citenamefont
		{Fu}, \citenamefont {Joannopoulos},\ and\ \citenamefont
		{Soljačić}}]{Lu2015}%
	\BibitemOpen
	\bibfield  {author} {\bibinfo {author} {\bibfnamefont {Ling}\ \bibnamefont
			{Lu}}, \bibinfo {author} {\bibfnamefont {Zhiyu}\ \bibnamefont {Wang}},
		\bibinfo {author} {\bibfnamefont {Dexin}\ \bibnamefont {Ye}}, \bibinfo
		{author} {\bibfnamefont {Lixin}\ \bibnamefont {Ran}}, \bibinfo {author}
		{\bibfnamefont {Liang}\ \bibnamefont {Fu}}, \bibinfo {author} {\bibfnamefont
			{John~D.}\ \bibnamefont {Joannopoulos}}, \ and\ \bibinfo {author}
		{\bibfnamefont {Marin}\ \bibnamefont {Soljačić}},\ }\bibfield  {title}
	{\enquote {\bibinfo {title} {Experimental observation of weyl points},}\
	}\href {\doibase 10.1126/science.aaa9273} {\bibfield  {journal} {\bibinfo
			{journal} {Science}\ }\textbf {\bibinfo {volume} {349}},\ \bibinfo {pages}
		{622--624} (\bibinfo {year} {2015})}\BibitemShut {NoStop}%
	\bibitem [{\citenamefont {Leykam}\ \emph {et~al.}(2016)\citenamefont {Leykam},
		\citenamefont {Rechtsman},\ and\ \citenamefont {Chong}}]{Leykam2016}%
	\BibitemOpen
	\bibfield  {author} {\bibinfo {author} {\bibfnamefont {Daniel}\ \bibnamefont
			{Leykam}}, \bibinfo {author} {\bibfnamefont {M.C.}\ \bibnamefont
			{Rechtsman}}, \ and\ \bibinfo {author} {\bibfnamefont {Y.D.}\ \bibnamefont
			{Chong}},\ }\bibfield {title} {\enquote {\bibinfo {title} {Anomalous
				topological phases and unpaired dirac cones in photonic floquet topological
				insulators},}\ }\href {\doibase 10.1103/physrevlett.117.013902} {\bibfield
		{journal} {\bibinfo  {journal} {Physical Review Letters}\ }\textbf {\bibinfo
			{volume} {117}},\ \bibinfo {pages} {013902} (\bibinfo {year}
		{2016})}\BibitemShut {NoStop}%
	\bibitem [{\citenamefont {Chiu}\ \emph {et~al.}(2016)\citenamefont {Chiu},
		\citenamefont {Teo}, \citenamefont {Schnyder},\ and\ \citenamefont
		{Ryu}}]{Chiu2016}%
	\BibitemOpen
	\bibfield  {author} {\bibinfo {author} {\bibfnamefont {Ching-Kai}\
			\bibnamefont {Chiu}}, \bibinfo {author} {\bibfnamefont {Jeffrey~C.Y.}\
			\bibnamefont {Teo}}, \bibinfo {author} {\bibfnamefont {Andreas~P.}\
			\bibnamefont {Schnyder}}, \ and\ \bibinfo {author} {\bibfnamefont {Shinsei}\
			\bibnamefont {Ryu}},\ }\bibfield  {title} {\enquote {\bibinfo {title}
			{Classification of topological quantum matter with symmetries},}\ }\href
	{\doibase 10.1103/revmodphys.88.035005} {\bibfield  {journal} {\bibinfo
			{journal} {Reviews of Modern Physics}\ }\textbf {\bibinfo {volume} {88}},\
		\bibinfo {pages} {035005} (\bibinfo {year} {2016})}\BibitemShut {NoStop}%
	\bibitem [{\citenamefont {Kunst}\ \emph {et~al.}(2018)\citenamefont {Kunst},
		\citenamefont {van Miert},\ and\ \citenamefont {Bergholtz}}]{Kunst2018}%
	\BibitemOpen
	\bibfield  {author} {\bibinfo {author} {\bibfnamefont {Flore~K.}\
			\bibnamefont {Kunst}}, \bibinfo {author} {\bibfnamefont {Guido}\ \bibnamefont
			{van Miert}}, \ and\ \bibinfo {author} {\bibfnamefont {Emil~J.}\ \bibnamefont
			{Bergholtz}},\ }\bibfield  {title} {\enquote {\bibinfo {title} {Lattice
				models with exactly solvable topological hinge and corner states},}\ }\href
	{\doibase 10.1103/physrevb.97.241405} {\bibfield  {journal} {\bibinfo
			{journal} {Physical Review B}\ }\textbf {\bibinfo {volume} {97}},\ \bibinfo
		{pages} {241405} (\bibinfo {year} {2018})}\BibitemShut {NoStop}%
	\bibitem [{\citenamefont {Armitage}\ \emph {et~al.}(2018)\citenamefont
		{Armitage}, \citenamefont {Mele},\ and\ \citenamefont
		{Vishwanath}}]{Armitage2018}%
	\BibitemOpen
	\bibfield  {author} {\bibinfo {author} {\bibfnamefont {N.P.}\ \bibnamefont
			{Armitage}}, \bibinfo {author} {\bibfnamefont {E.J.}\ \bibnamefont
			{Mele}}, \ and\ \bibinfo {author} {\bibfnamefont {Ashvin}\ \bibnamefont
			{Vishwanath}},\ }\bibfield  {title} {\enquote {\bibinfo {title} {Weyl and
				dirac semimetals in three-dimensional solids},}\ }\href {\doibase
		10.1103/revmodphys.90.015001} {\bibfield  {journal} {\bibinfo  {journal}
			{Reviews of Modern Physics}\ }\textbf {\bibinfo {volume} {90}},\ \bibinfo
		{pages} {015001} (\bibinfo {year} {2018})}\BibitemShut {NoStop}%
	\bibitem [{\citenamefont {Su}\ \emph {et~al.}(1979)\citenamefont {Su},
		\citenamefont {Schrieffer},\ and\ \citenamefont {Heeger}}]{Su1979}%
	\BibitemOpen
	\bibfield  {author} {\bibinfo {author} {\bibfnamefont {W.~P.}\ \bibnamefont
			{Su}}, \bibinfo {author} {\bibfnamefont {J.~R.}\ \bibnamefont {Schrieffer}},
		\ and\ \bibinfo {author} {\bibfnamefont {A.~J.}\ \bibnamefont {Heeger}},\
	}\bibfield  {title} {\enquote {\bibinfo {title} {Solitons in
				polyacetylene},}\ }\href {\doibase 10.1103/physrevlett.42.1698} {\bibfield
		{journal} {\bibinfo  {journal} {Physical Review Letters}\ }\textbf {\bibinfo
			{volume} {42}},\ \bibinfo {pages} {1698--1701} (\bibinfo {year}
		{1979})}\BibitemShut {NoStop}%
	\bibitem [{\citenamefont {Zak}(1989)}]{Zak1989}%
	\BibitemOpen
	\bibfield  {author} {\bibinfo {author} {\bibfnamefont {J.}~\bibnamefont
			{Zak}},\ }\bibfield  {title} {\enquote {\bibinfo {title} {Berry’s phase for
				energy bands in solids},}\ }\href {\doibase 10.1103/physrevlett.62.2747}
	{\bibfield  {journal} {\bibinfo  {journal} {Physical Review Letters}\
		}\textbf {\bibinfo {volume} {62}},\ \bibinfo {pages} {2747--2750} (\bibinfo
		{year} {1989})}\BibitemShut {NoStop}%
	\bibitem [{\citenamefont {Stepanenko}\ and\ \citenamefont
		{Gorlach}(2020)}]{Stepanenko2020}%
	\BibitemOpen
	\bibfield  {author} {\bibinfo {author} {\bibfnamefont {Andrei~A.}\
			\bibnamefont {Stepanenko}}\ and\ \bibinfo {author} {\bibfnamefont {Maxim~A.}\
			\bibnamefont {Gorlach}},\ }\bibfield  {title} {\enquote {\bibinfo {title}
			{Interaction-induced topological states of photon pairs},}\ }\href {\doibase
		10.1103/physreva.102.013510} {\bibfield  {journal} {\bibinfo  {journal}
			{Physical Review A}\ }\textbf {\bibinfo {volume} {102}},\ \bibinfo {pages}
		{013510} (\bibinfo {year} {2020})}\BibitemShut {NoStop}%
	\bibitem [{\citenamefont {Winkler}\ \emph {et~al.}(2006)\citenamefont
		{Winkler}, \citenamefont {Thalhammer}, \citenamefont {Lang}, \citenamefont
		{Grimm}, \citenamefont {Hecker~Denschlag}, \citenamefont {Daley},
		\citenamefont {Kantian}, \citenamefont {Büchler},\ and\ \citenamefont
		{Zoller}}]{Winkler2006}%
	\BibitemOpen
	\bibfield  {author} {\bibinfo {author} {\bibfnamefont {K.}~\bibnamefont
			{Winkler}}, \bibinfo {author} {\bibfnamefont {G.}~\bibnamefont {Thalhammer}},
		\bibinfo {author} {\bibfnamefont {F.}~\bibnamefont {Lang}}, \bibinfo {author}
		{\bibfnamefont {R.}~\bibnamefont {Grimm}}, \bibinfo {author} {\bibfnamefont
			{J.}~\bibnamefont {Hecker~Denschlag}}, \bibinfo {author} {\bibfnamefont
			{A.~J.}\ \bibnamefont {Daley}}, \bibinfo {author} {\bibfnamefont
			{A.}~\bibnamefont {Kantian}}, \bibinfo {author} {\bibfnamefont {H.~P.}\
			\bibnamefont {Büchler}}, \ and\ \bibinfo {author} {\bibfnamefont
			{P.}~\bibnamefont {Zoller}},\ }\bibfield  {title} {\enquote {\bibinfo {title}
			{Repulsively bound atom pairs in an optical lattice},}\ }\href {\doibase
		10.1038/nature04918} {\bibfield  {journal} {\bibinfo  {journal} {Nature}\
		}\textbf {\bibinfo {volume} {441}},\ \bibinfo {pages} {853--856} (\bibinfo
		{year} {2006})}\BibitemShut {NoStop}%
	\bibitem [{\citenamefont {Fölling}\ \emph {et~al.}(2007)\citenamefont
		{Fölling}, \citenamefont {Trotzky}, \citenamefont {Cheinet}, \citenamefont
		{Feld}, \citenamefont {Saers}, \citenamefont {Widera}, \citenamefont
		{Müller},\ and\ \citenamefont {Bloch}}]{Foelling2007}%
	\BibitemOpen
	\bibfield  {author} {\bibinfo {author} {\bibfnamefont {S.}~\bibnamefont
			{Fölling}}, \bibinfo {author} {\bibfnamefont {S.}~\bibnamefont {Trotzky}},
		\bibinfo {author} {\bibfnamefont {P.}~\bibnamefont {Cheinet}}, \bibinfo
		{author} {\bibfnamefont {M.}~\bibnamefont {Feld}}, \bibinfo {author}
		{\bibfnamefont {R.}~\bibnamefont {Saers}}, \bibinfo {author} {\bibfnamefont
			{A.}~\bibnamefont {Widera}}, \bibinfo {author} {\bibfnamefont
			{T.}~\bibnamefont {Müller}}, \ and\ \bibinfo {author} {\bibfnamefont
			{I.}~\bibnamefont {Bloch}},\ }\bibfield  {title} {\enquote {\bibinfo {title}
			{Direct observation of second-order atom tunnelling},}\ }\href {\doibase
		10.1038/nature06112} {\bibfield  {journal} {\bibinfo  {journal} {Nature}\
		}\textbf {\bibinfo {volume} {448}},\ \bibinfo {pages} {1029--1032} (\bibinfo
		{year} {2007})}\BibitemShut {NoStop}%
	\bibitem [{\citenamefont {Gustavsson}\ \emph {et~al.}(2008)\citenamefont
		{Gustavsson}, \citenamefont {Haller}, \citenamefont {Mark}, \citenamefont
		{Danzl}, \citenamefont {Rojas-Kopeinig},\ and\ \citenamefont
		{Nägerl}}]{Gustavsson2008}%
	\BibitemOpen
	\bibfield  {author} {\bibinfo {author} {\bibfnamefont {M.}~\bibnamefont
			{Gustavsson}}, \bibinfo {author} {\bibfnamefont {E.}~\bibnamefont {Haller}},
		\bibinfo {author} {\bibfnamefont {M.~J.}\ \bibnamefont {Mark}}, \bibinfo
		{author} {\bibfnamefont {J.~G.}\ \bibnamefont {Danzl}}, \bibinfo {author}
		{\bibfnamefont {G.}~\bibnamefont {Rojas-Kopeinig}}, \ and\ \bibinfo {author}
		{\bibfnamefont {H.-C.}\ \bibnamefont {Nägerl}},\ }\bibfield  {title}
	{\enquote {\bibinfo {title} {Control of interaction-induced dephasing of
				bloch oscillations},}\ }\href {\doibase 10.1103/physrevlett.100.080404}
	{\bibfield  {journal} {\bibinfo  {journal} {Physical Review Letters}\
		}\textbf {\bibinfo {volume} {100}},\ \bibinfo {pages} {080404} (\bibinfo
		{year} {2008})}\BibitemShut {NoStop}%
	\bibitem [{\citenamefont {Mahajan}\ and\ \citenamefont
		{Thyagaraja}(2006)}]{Mahajan2006}%
	\BibitemOpen
	\bibfield  {author} {\bibinfo {author} {\bibfnamefont {S~M}\ \bibnamefont
			{Mahajan}}\ and\ \bibinfo {author} {\bibfnamefont {A}~\bibnamefont
			{Thyagaraja}},\ }\bibfield  {title} {\enquote {\bibinfo {title} {Exact
				two-body bound states with coulomb repulsion in a periodic potential},}\
	}\href {\doibase 10.1088/0305-4470/39/47/l01} {\bibfield  {journal} {\bibinfo
			{journal} {Journal of Physics A: Mathematical and General}\ }\textbf
		{\bibinfo {volume} {39}},\ \bibinfo {pages} {L667--L671} (\bibinfo {year}
		{2006})}\BibitemShut {NoStop}%
	\bibitem [{\citenamefont {Petrosyan}\ \emph {et~al.}(2007)\citenamefont
		{Petrosyan}, \citenamefont {Schmidt}, \citenamefont {Anglin},\ and\
		\citenamefont {Fleischhauer}}]{Petrosyan2007}%
	\BibitemOpen
	\bibfield  {author} {\bibinfo {author} {\bibfnamefont {David}\ \bibnamefont
			{Petrosyan}}, \bibinfo {author} {\bibfnamefont {Bernd}\ \bibnamefont
			{Schmidt}}, \bibinfo {author} {\bibfnamefont {James~R.}\ \bibnamefont
			{Anglin}}, \ and\ \bibinfo {author} {\bibfnamefont {Michael}\ \bibnamefont
			{Fleischhauer}},\ }\bibfield  {title} {\enquote {\bibinfo {title} {Quantum
				liquid of repulsively bound pairs of particles in a lattice},}\ }\href
	{\doibase 10.1103/physreva.76.033606} {\bibfield  {journal} {\bibinfo
			{journal} {Physical Review A}\ }\textbf {\bibinfo {volume} {76}},\ \bibinfo
		{pages} {033606} (\bibinfo {year} {2007})}\BibitemShut {NoStop}%
	\bibitem [{\citenamefont {Creffield}(2007)}]{Creffield2007}%
	\BibitemOpen
	\bibfield  {author} {\bibinfo {author} {\bibfnamefont {C.~E.}\ \bibnamefont
			{Creffield}},\ }\bibfield  {title} {\enquote {\bibinfo {title} {Coherent
				control of self-trapping of cold bosonic atoms},}\ }\href {\doibase
		10.1103/physreva.75.031607} {\bibfield  {journal} {\bibinfo  {journal}
			{Physical Review A}\ }\textbf {\bibinfo {volume} {75}},\ \bibinfo {pages}
		{031607} (\bibinfo {year} {2007})}\BibitemShut {NoStop}%
	\bibitem [{\citenamefont {Kuklov}\ and\ \citenamefont
		{Moritz}(2007)}]{Kuklov2007}%
	\BibitemOpen
	\bibfield  {author} {\bibinfo {author} {\bibfnamefont {Anatoly}\ \bibnamefont
			{Kuklov}}\ and\ \bibinfo {author} {\bibfnamefont {Henning}\ \bibnamefont
			{Moritz}},\ }\bibfield  {title} {\enquote {\bibinfo {title} {Detecting
				multiatomic composite states in optical lattices},}\ }\href {\doibase
		10.1103/physreva.75.013616} {\bibfield  {journal} {\bibinfo  {journal}
			{Physical Review A}\ }\textbf {\bibinfo {volume} {75}},\ \bibinfo {pages}
		{013616} (\bibinfo {year} {2007})}\BibitemShut {NoStop}%
	\bibitem [{\citenamefont {Zöllner}\ \emph {et~al.}(2008)\citenamefont
		{Zöllner}, \citenamefont {Meyer},\ and\ \citenamefont
		{Schmelcher}}]{Zoellner2008}%
	\BibitemOpen
	\bibfield  {author} {\bibinfo {author} {\bibfnamefont {Sascha}\ \bibnamefont
			{Zöllner}}, \bibinfo {author} {\bibfnamefont {Hans-Dieter}\ \bibnamefont
			{Meyer}}, \ and\ \bibinfo {author} {\bibfnamefont {Peter}\ \bibnamefont
			{Schmelcher}},\ }\bibfield  {title} {\enquote {\bibinfo {title} {Few-boson
				dynamics in double wells: From single-atom to correlated pair tunneling},}\
	}\href {\doibase 10.1103/physrevlett.100.040401} {\bibfield  {journal}
		{\bibinfo  {journal} {Physical Review Letters}\ }\textbf {\bibinfo {volume}
			{100}},\ \bibinfo {pages} {040401} (\bibinfo {year} {2008})}\BibitemShut
	{NoStop}%
	\bibitem [{\citenamefont {Wang}\ \emph {et~al.}(2008)\citenamefont {Wang},
		\citenamefont {Hao},\ and\ \citenamefont {Chen}}]{Wang2008}%
	\BibitemOpen
	\bibfield  {author} {\bibinfo {author} {\bibfnamefont {L.}~\bibnamefont
			{Wang}}, \bibinfo {author} {\bibfnamefont {Y.}~\bibnamefont {Hao}}, \ and\
		\bibinfo {author} {\bibfnamefont {S.}~\bibnamefont {Chen}},\ }\bibfield
	{title} {\enquote {\bibinfo {title} {Quantum dynamics of repulsively bound
				atom pairs in the bose-hubbard model},}\ }\href {\doibase
		10.1140/epjd/e2008-00077-3} {\bibfield  {journal} {\bibinfo  {journal} {The
				European Physical Journal D}\ }\textbf {\bibinfo {volume} {48}},\ \bibinfo
		{pages} {229--234} (\bibinfo {year} {2008})}\BibitemShut {NoStop}%
	\bibitem [{\citenamefont {Valiente}\ and\ \citenamefont
		{Petrosyan}(2008)}]{Valiente2008}%
	\BibitemOpen
	\bibfield  {author} {\bibinfo {author} {\bibfnamefont {M}~\bibnamefont
			{Valiente}}\ and\ \bibinfo {author} {\bibfnamefont {D}~\bibnamefont
			{Petrosyan}},\ }\bibfield  {title} {\enquote {\bibinfo {title} {Two-particle
				states in the hubbard model},}\ }\href {\doibase
		10.1088/0953-4075/41/16/161002} {\bibfield  {journal} {\bibinfo  {journal}
			{Journal of Physics B: Atomic, Molecular and Optical Physics}\ }\textbf
		{\bibinfo {volume} {41}},\ \bibinfo {pages} {161002} (\bibinfo {year}
		{2008})}\BibitemShut {NoStop}%
	\bibitem [{\citenamefont {Jin}\ \emph {et~al.}(2009)\citenamefont {Jin},
		\citenamefont {Chen},\ and\ \citenamefont {Song}}]{Jin2009}%
	\BibitemOpen
	\bibfield  {author} {\bibinfo {author} {\bibfnamefont {L.}~\bibnamefont
			{Jin}}, \bibinfo {author} {\bibfnamefont {B.}~\bibnamefont {Chen}}, \ and\
		\bibinfo {author} {\bibfnamefont {Z.}~\bibnamefont {Song}},\ }\bibfield
	{title} {\enquote {\bibinfo {title} {Coherent shift of localized bound pairs
				in the bose-hubbard model},}\ }\href {\doibase 10.1103/physreva.79.032108}
	{\bibfield  {journal} {\bibinfo  {journal} {Physical Review A}\ }\textbf
		{\bibinfo {volume} {79}},\ \bibinfo {pages} {032108} (\bibinfo {year}
		{2009})}\BibitemShut {NoStop}%
	\bibitem [{\citenamefont {Valiente}\ and\ \citenamefont
		{Petrosyan}(2009)}]{Valiente2009}%
	\BibitemOpen
	\bibfield  {author} {\bibinfo {author} {\bibfnamefont {M}~\bibnamefont
			{Valiente}}\ and\ \bibinfo {author} {\bibfnamefont {D}~\bibnamefont
			{Petrosyan}},\ }\bibfield  {title} {\enquote {\bibinfo {title} {Scattering
				resonances and two-particle bound states of the extended hubbard model},}\
	}\href {\doibase 10.1088/0953-4075/42/12/121001} {\bibfield  {journal}
		{\bibinfo  {journal} {Journal of Physics B: Atomic, Molecular and Optical
				Physics}\ }\textbf {\bibinfo {volume} {42}},\ \bibinfo {pages} {121001}
		(\bibinfo {year} {2009})}\BibitemShut {NoStop}%
	\bibitem [{\citenamefont {Valiente}\ \emph {et~al.}(2010)\citenamefont
		{Valiente}, \citenamefont {Petrosyan},\ and\ \citenamefont
		{Saenz}}]{Valiente2010}%
	\BibitemOpen
	\bibfield  {author} {\bibinfo {author} {\bibfnamefont {Manuel}\ \bibnamefont
			{Valiente}}, \bibinfo {author} {\bibfnamefont {David}\ \bibnamefont
			{Petrosyan}}, \ and\ \bibinfo {author} {\bibfnamefont {Alejandro}\
			\bibnamefont {Saenz}},\ }\bibfield  {title} {\enquote {\bibinfo {title}
			{Three-body bound states in a lattice},}\ }\href {\doibase
		10.1103/physreva.81.011601} {\bibfield  {journal} {\bibinfo  {journal}
			{Physical Review A}\ }\textbf {\bibinfo {volume} {81}},\ \bibinfo {pages}
		{011601} (\bibinfo {year} {2010})}\BibitemShut {NoStop}%
	\bibitem [{\citenamefont {Javanainen}\ \emph {et~al.}(2010)\citenamefont
		{Javanainen}, \citenamefont {Odong},\ and\ \citenamefont
		{Sanders}}]{Javanainen2010}%
	\BibitemOpen
	\bibfield  {author} {\bibinfo {author} {\bibfnamefont {Juha}\ \bibnamefont
			{Javanainen}}, \bibinfo {author} {\bibfnamefont {Otim}\ \bibnamefont
			{Odong}}, \ and\ \bibinfo {author} {\bibfnamefont {Jerome~C.}\ \bibnamefont
			{Sanders}},\ }\bibfield  {title} {\enquote {\bibinfo {title} {Dimer of two
				bosons in a one-dimensional optical lattice},}\ }\href {\doibase
		10.1103/physreva.81.043609} {\bibfield  {journal} {\bibinfo  {journal}
			{Physical Review A}\ }\textbf {\bibinfo {volume} {81}},\ \bibinfo {pages}
		{043609} (\bibinfo {year} {2010})}\BibitemShut {NoStop}%
	\bibitem [{\citenamefont {Wang}\ and\ \citenamefont {Liang}(2010)}]{Wang2010}%
	\BibitemOpen
	\bibfield  {author} {\bibinfo {author} {\bibfnamefont {Y.-M.}\ \bibnamefont
			{Wang}}\ and\ \bibinfo {author} {\bibfnamefont {J.-Q.}\ \bibnamefont
			{Liang}},\ }\bibfield  {title} {\enquote {\bibinfo {title} {Repulsive
				bound-atom pairs in an optical lattice with two-body interaction of nearest
				neighbors},}\ }\href {\doibase 10.1103/physreva.81.045601} {\bibfield
		{journal} {\bibinfo  {journal} {Physical Review A}\ }\textbf {\bibinfo
			{volume} {81}},\ \bibinfo {pages} {045601} (\bibinfo {year}
		{2010})}\BibitemShut {NoStop}%
	\bibitem [{\citenamefont {Rosch}\ \emph {et~al.}(2008)\citenamefont {Rosch},
		\citenamefont {Rasch}, \citenamefont {Binz},\ and\ \citenamefont
		{Vojta}}]{Rosch2008}%
	\BibitemOpen
	\bibfield  {author} {\bibinfo {author} {\bibfnamefont {Achim}\ \bibnamefont
			{Rosch}}, \bibinfo {author} {\bibfnamefont {David}\ \bibnamefont {Rasch}},
		\bibinfo {author} {\bibfnamefont {Benedikt}\ \bibnamefont {Binz}}, \ and\
		\bibinfo {author} {\bibfnamefont {Matthias}\ \bibnamefont {Vojta}},\
	}\bibfield  {title} {\enquote {\bibinfo {title} {Metastable superfluidity of
				repulsive fermionic atoms in optical lattices},}\ }\href {\doibase
		10.1103/physrevlett.101.265301} {\bibfield  {journal} {\bibinfo  {journal}
			{Physical Review Letters}\ }\textbf {\bibinfo {volume} {101}},\ \bibinfo
		{pages} {265301} (\bibinfo {year} {2008})}\BibitemShut {NoStop}%
	\bibitem [{\citenamefont {Zhang}(2024)}]{Zhang2024}%
	\BibitemOpen
	\bibfield  {author} {\bibinfo {author} {\bibfnamefont {Kun-Liang}\
			\bibnamefont {Zhang}},\ }\bibfield  {title} {\enquote {\bibinfo {title}
			{Doublons bloch oscillations in the mass-imbalanced extended fermi-hubbard
				model},}\ }\href {\doibase 10.48550/ARXIV.2408.08194} {\  (\bibinfo {year}
		{2024}),\ 10.48550/ARXIV.2408.08194},\ \Eprint
	{http://arxiv.org/abs/2408.08194} {arXiv:2408.08194 [cond-mat.quant-gas]}
	\BibitemShut {NoStop}%
	\bibitem [{\citenamefont {Jin}\ and\ \citenamefont {Song}(2011)}]{Jin2011}%
	\BibitemOpen
	\bibfield  {author} {\bibinfo {author} {\bibfnamefont {L}~\bibnamefont
			{Jin}}\ and\ \bibinfo {author} {\bibfnamefont {Z}~\bibnamefont {Song}},\
	}\bibfield  {title} {\enquote {\bibinfo {title} {Fast transfer and efficient
				coherent separation of a bound cluster in the extended hubbard model},}\
	}\href {\doibase 10.1088/1367-2630/13/6/063009} {\bibfield  {journal}
		{\bibinfo  {journal} {New Journal of Physics}\ }\textbf {\bibinfo {volume}
			{13}},\ \bibinfo {pages} {063009} (\bibinfo {year} {2011})}\BibitemShut
	{NoStop}%
	\bibitem [{\citenamefont {Rice}\ and\ \citenamefont {Mele}(1982)}]{Rice1982}%
	\BibitemOpen
	\bibfield  {author} {\bibinfo {author} {\bibfnamefont {M.~J.}\ \bibnamefont
			{Rice}}\ and\ \bibinfo {author} {\bibfnamefont {E.~J.}\ \bibnamefont
			{Mele}},\ }\bibfield  {title} {\enquote {\bibinfo {title} {Elementary
				excitations of a linearly conjugated diatomic polymer},}\ }\href {\doibase
		10.1103/physrevlett.49.1455} {\bibfield  {journal} {\bibinfo  {journal}
			{Physical Review Letters}\ }\textbf {\bibinfo {volume} {49}},\ \bibinfo
		{pages} {1455--1459} (\bibinfo {year} {1982})}\BibitemShut {NoStop}%
	\bibitem [{\citenamefont {Xiao}\ \emph {et~al.}(2010)\citenamefont {Xiao},
		\citenamefont {Chang},\ and\ \citenamefont {Niu}}]{Xiao2010}%
	\BibitemOpen
	\bibfield  {author} {\bibinfo {author} {\bibfnamefont {Di}~\bibnamefont
			{Xiao}}, \bibinfo {author} {\bibfnamefont {Ming-Che}\ \bibnamefont {Chang}},
		\ and\ \bibinfo {author} {\bibfnamefont {Qian}\ \bibnamefont {Niu}},\
	}\bibfield  {title} {\enquote {\bibinfo {title} {Berry phase effects on
				electronic properties},}\ }\href {\doibase 10.1103/revmodphys.82.1959}
	{\bibfield  {journal} {\bibinfo  {journal} {Reviews of Modern Physics}\
		}\textbf {\bibinfo {volume} {82}},\ \bibinfo {pages} {1959--2007} (\bibinfo
		{year} {2010})}\BibitemShut {NoStop}%
	\bibitem [{\citenamefont {Wang}\ \emph
		{et~al.}(2018{\natexlab{a}})\citenamefont {Wang}, \citenamefont {Li},
		\citenamefont {Zhang},\ and\ \citenamefont {Song}}]{Wang2018}%
	\BibitemOpen
	\bibfield  {author} {\bibinfo {author} {\bibfnamefont {R.}~\bibnamefont
			{Wang}}, \bibinfo {author} {\bibfnamefont {C.}~\bibnamefont {Li}}, \bibinfo
		{author} {\bibfnamefont {X.~Z.}\ \bibnamefont {Zhang}}, \ and\ \bibinfo
		{author} {\bibfnamefont {Z.}~\bibnamefont {Song}},\ }\bibfield  {title}
	{\enquote {\bibinfo {title} {Dynamical bulk-edge correspondence for
				degeneracy lines in parameter space},}\ }\href {\doibase
		10.1103/physrevb.98.014303} {\bibfield  {journal} {\bibinfo  {journal}
			{Physical Review B}\ }\textbf {\bibinfo {volume} {98}},\ \bibinfo {pages}
		{014303} (\bibinfo {year} {2018}{\natexlab{a}})}\BibitemShut {NoStop}%
	\bibitem [{\citenamefont {Wang}\ \emph
		{et~al.}(2018{\natexlab{b}})\citenamefont {Wang}, \citenamefont {Zhang},\
		and\ \citenamefont {Song}}]{Wang2018a}%
	\BibitemOpen
	\bibfield  {author} {\bibinfo {author} {\bibfnamefont {R.}~\bibnamefont
			{Wang}}, \bibinfo {author} {\bibfnamefont {X.~Z.}\ \bibnamefont {Zhang}}, \
		and\ \bibinfo {author} {\bibfnamefont {Z.}~\bibnamefont {Song}},\ }\bibfield
	{title} {\enquote {\bibinfo {title} {Dynamical topological invariant for the
				non-hermitian rice-mele model},}\ }\href {\doibase
		10.1103/physreva.98.042120} {\bibfield  {journal} {\bibinfo  {journal}
			{Physical Review A}\ }\textbf {\bibinfo {volume} {98}},\ \bibinfo {pages}
		{042120} (\bibinfo {year} {2018}{\natexlab{b}})}\BibitemShut {NoStop}%
\end{thebibliography}
\end{document}